\documentclass[12pt]{iopart}

\expandafter\let\csname equation*\endcsname\relax
\expandafter\let\csname endequation*\endcsname\relax

\usepackage{
    amsfonts,
    amsmath,
    amssymb,
    amsthm,
    bm,
    braket,
    dsfont,
    graphicx,
    latexsym,
    mathpazo,
    mathrsfs, 
}

\usepackage[hidelinks]{hyperref}
\usepackage{bm}

\usepackage{tikz}
\usetikzlibrary{quantikz2}

\usepackage{amssymb,amscd}
\usepackage{todonotes}
\usepackage{graphicx}
\usepackage{braket}
\usepackage{dsfont}
\usepackage{color}
\usepackage{tikz}
\usetikzlibrary{arrows,positioning}
\usetikzlibrary{shapes.geometric}
\usetikzlibrary{patterns}
\usepackage{mathrsfs}
\usepackage{algorithm}
\usepackage{algpseudocode}
\algrenewcommand\algorithmicrequire{\textbf{Input:}}
\algrenewcommand\algorithmicensure{\textbf{Output:}}
\usepackage{mathtools}
\usepackage{tcolorbox}
\usepackage[utf8]{inputenc}
\usepackage{enumerate}
\usepackage[english]{babel}
\usepackage{lipsum} 

\usepackage{appendix}

\newcommand\kket[1]{\vert#1\rangle\!\rangle}
\newcommand\bbra[1]{\langle\!\langle#1\vert}

\newcommand{\groupname}{rHDG}
\newcommand{\myi}{\ensuremath{\mathrm{i}}}

\usepackage{nicefrac}
\newcommand\dyad[1]{\vert#1\rangle\!\langle#1\vert}
\DeclareMathOperator{\trace}{tr}
\newcommand\abs[1]{\vert#1\vert}
\usepackage{bm}

\numberwithin{equation}{section}
\DeclareMathOperator{\Range}{Image}
\DeclareMathOperator{\Diag}{diag}
\DeclareMathOperator{\toRing}{Ring}
\DeclareMathOperator{\Span}{span}
\DeclareMathOperator{\Order}{\mathfrak{o}}
\DeclareMathOperator{\Ring}{\mathbb{Z}}
\DeclareMathOperator{\Fidelity}{\bm{F}}
\DeclareMathOperator*{\Average}{\mathbb{E}}
\DeclareMathOperator*{\average}{\mathbb{E}}
\DeclareMathOperator{\RepCyclic}{\mathsf{D}}
\DeclareMathOperator{\RepSymm}{\mathsf{S}}
\newcommand{\Eigenvalue}{\eta}
\newcommand{\Character}{\chi}
\newcommand{\Irrep}{\Sigma}
\newcommand{\Permutation}{\sigma}
\newcommand{\QuditDim}{d}
\newcommand{\qudim}{d}

\newcommand{\startnumbering}{0}
\newcommand{\RootUnity}{\omega}
\newcommand{\Partition}{\lambda}
\newcommand{\CyclicElement}{\alpha}
\newcommand{\UniRepHdg}{\gamma}
\newcommand{\RepPlRep}{\Gamma}
\newcommand{\Projector}{\Pi}
\newcommand{\IrrepIndex}{\varpi}
\newcommand{\somestate}{\psi}
\newcommand{\Channel}{\mathcal{E}}
\newcommand{\cliffordset}{\mathcal{C}}
\newcommand{\NullVector}{\bm{0}}
\newcommand{\SetOfChannels}{\mathscr{B}}
\newcommand{\FiniteGroup}{\mathbb{G}}
\newcommand{\SetOfMatrices}{\mathscr{M}}
\newcommand{\DensityMatrices}{\mathscr{D}}
\newcommand{\DensityMatrix}{\rho}
\newcommand{\GroupElement}{g}

\newcommand{\SetOfKraus}{\bm{A}}
\newcommand{\TwirlOp}{\mathcal{T}}
\newcommand{\CircuitDepth}{m}

\newcommand{\identity}{\mathbb{I}}
\newcommand{\channel}{\mathcal{E}}
\newcommand{\numberOfQudits}{n}
\newcommand{\integers}{\mathbb{Z}^{+}}

\usepackage{xspace}
\newcommand\thegroup{\ensuremath{{}_d\mathrm{rHDG}}}
\newcommand{\labelirrepstd}{\mathrm{std}}

\usepackage{amsthm}

\usepackage{thmtools}
\usepackage{thm-restate}
\usepackage{cleveref}
\declaretheorem[name=Theorem,numberwithin=section]{theorem}
\newtheorem{lemma}[theorem]{Lemma} 
\newtheorem{corollary}[theorem]{Corollary}

\theoremstyle{definition}

\newcommand{\numberlements}{r}
\newcommand{\setelement}{q}
\newcommand{\finiteset}{\bm{q}}
\newcommand{\auxdummyindex}{l}
\newcommand{\orderof}{\mathfrak{o}}
\newcommand{\rootunit}{\omega}

\usepackage{multirow}

\usepackage[inline]{enumitem}

\usepackage[normalem]{ulem}
\newcommand\redsout{\bgroup\markoverwith{\textcolor{red}{\rule[0.5ex]{2pt}{0.4pt}}}\ULon}

\begin{document}

\title{Randomised benchmarking for universal qudit gates}
\author{David Amaro-Alcal\'a\footnote{Corresponding author.}}
\ead{david.amaroalcala@ucalgary.ca}
\address{Institute for Quantum Science and Technology, University of Calgary,
 Alberta T2N~1N4, Canada}
 \address{Department of Physics, Lakehead University, Thunder Bay, ON, P7B 5E1}
\author{Barry C. Sanders}
\ead{sandersb@ucalgary.ca}
\address{Institute for Quantum Science and Technology, University of Calgary,
 Alberta T2N~1N4, Canada}
\author{Hubert de~Guise}%
\ead{hubert.deguise@lakeheadu.ca}
 \address{Department of Physics, Lakehead University, Thunder Bay, ON, P7B 5E1}
\address{Institute for Quantum Science and Technology, University of Calgary,
 Alberta T2N~1N4, Canada}
\date{\today}%

\begin{abstract}
We aim to establish a scalable scheme for characterising 
diagonal non-Clifford gates
for single- and multi-qudit systems; \(d\) is a prime-power integer.
By employing cyclic operators and a qudit T gate,
we generalise the dihedral benchmarking scheme for single- and multi-qudit
circuits.
Our results establish a path for experimentally benchmarking qudit systems and
are of theoretical and experimental interest because 
our scheme is optimal insofar as it does not require preparation of the full qudit Clifford
gate set to characterise a non-Clifford gate.
Moreover, combined with Clifford randomised benchmarking, 
our scheme is useful to characterise the generators of a universal gate set.
\end{abstract}

\section{Introduction}
Driven by the need to exploit the entire Hilbert space available in Physics
experiments,
the extension from qubit information processing to qudits is now experimentally feasible.
The main motivation for choosing qudits over qubits is to avoid truncating
naturally higher dimensional quantum
systems into qubits~\cite{brylinski2002universal,Wang2020}.
Single- and multi-qudit experiments use
photons~\cite{Meng2024,Chi2022,Imany2019,Lanyon2009,min2018,Erhard2017,Liu2020,Liu2020b,Malik2016,Lanyon_Weinhold_Langford_OBrien_Resch_Gilchrist_White_2008}, 
trapped
ions~\cite{randall2015,leupold18,Klimov_Guzman_Retamal_Saavedra_2003,Zhang2013,
Hrmo2023,Aksenov2023},
superconducting
qudits~\cite{Luo2023,Seifert2023,Roy2022,Kononenko_Yurtalan_Ren_Shi_Ashhab_Lupascu_2021,
Morvan_Ramasesh_Blok_Kreikebaum_OBrien_Chen_Mitchell_Naik_Santiago_Siddiqi_2021,Goss2022}, 
dopants in silicon~\cite{fernandez2022coherent}, ultracold atoms~\cite{Lindon2022}, 
and spin systems~\cite{Fu2022,guo2023}.
Qudits 
systems are currently in use in
quantum communication~\cite{cozzolino2019},
quantum teleportation~\cite{Luo_Zhong_Erhard_Wang_Peng_Krenn_Jiang_Li_Liu_Lu_2019, Hu_Zhang_Liu_Cai_Ye_Guo_Xing_Huang_Huang_Li_2020},
quantum memories~\cite{Vashukevich_Bashmakova_Golubeva_Golubev_2022,Otten_Kapoor_Ozguler_Holland_Kowalkowski_Alexeev_Lyon_2021},
Bell-state measurements~\cite{Zhang_Zhang_Hu_Liu_Huang_Li_Guo_2019}, 
study of spin chains~\cite{Senko_Richerme_Smith_Lee_Cohen_Retzker_Monroe_2015,Blok_Ramasesh_Schuster_OBrien_Kreikebaum_Dahlen_Morvan_Yoshida_Yao_Siddiqi_2021,Hu_Zhang_Liu_Cai_Ye_Guo_Xing_Huang_Huang_Li_2020,Imany2019,Zhang_Zhang_Hu_Liu_Huang_Li_Guo_2019,Lanyon_Weinhold_Langford_OBrien_Resch_Gilchrist_White_2008},
in enhancing quantum error correction techniques~\cite{Campbell2012,Campbell2014},
in encoding qubits~\cite{Katip2016} and qudits~\cite{Majumdar2018},
simulations of many-body systems~\cite{Fauseweh2023},
quantum key
distribution~\cite{bouchard2018,Nape18,Stasiuk2023},
simulation of high-energy physics~\cite{Holland2020,GonzlezCuadra2022,Bender2018,Gustafson2021,Zohar2013,Turro2022,Kurkcuoglu2021-tn,Bauer2023a,Bauer2023b},
and quantum computing~\cite{Chi2022,Nikolaeva2022,Luo_Wang_2014}.
In particular universal gate sets are necessary 
to perform quantum computing tasks currently unviable for classical machines.
Our work is about benchmarking qudits gates to assess their performance:
we extend randomised benchmarking (RB) to estimate the average gate fidelity of
single-qudit T gates and controlled-T gates.
With our extension of randomised benchmarking
to (characterise) T~gates,
the scalable characterisation of generators of a universal qudit gate set is
now available.

Randomised benchmarking is the standard for characterising qubit gate
performance~\cite{Emerson_Alicki_Zyczkowski_2005,Knill2008,dugas2015,Cross2016,Harper_Flammia_2017};
as a black box, a randomised benchmarking schemes receives as input a set of
gates and returns the fidelity of the gates.
However, the characterisation of single- and multi-qudit T gates has been
lacking~\cite{amaro2024,dugas2015,Cross2016,
Harper_Flammia_2017,Garion2020},
thereby preventing development of universal qudit RB,
a deficit that we fix here.
In generalising the widely used RB method,
we highlight here three important features, namely: 
(i)~straightforward circuit-design cost,
(ii)~robustness against state preparation and measurement (SPAM) errors,
and (iii)~feasibility.
Here, we introduce a scalable and optimal scheme---optimal with respect to the
required primitive gates (the generators of the gate set that
  are also
native gates of the platform): 
two cyclic gates, a~T gate, and a
Hadamard (H) gate---to characterise universal
single- and multi-qudit gates.
Note that the gate set we introduce does not generate a universal gate set:
the complete characterisation of a generating set of a universal gate set
requires characterising a Fourier matrix;
the characterisation of the Fourier matrix
can be done with standard Clifford randomised
benchmarking~\cite{Jafarzadeh_Wu_Sanders_Sanders_2020}.
Thus, concatenating our scheme with Clifford randomised benchmarking
characterises a universal gate set. 

Randomised benchmarking in general is concerned with estimating the average gate fidelity (AGF) over a gate set~\cite{Emerson_Alicki_Zyczkowski_2005,MagesanEaswar2012Emoq}. 
This estimate is computed by fitting an exponential decay curve to a
plot of sequence fidelity \emph{vs} circuit depth.
This sequence fidelity corresponds
to the state fidelity~\cite{MikeAndIke} between~\(\ket{0}\)
and a final state~\(\DensityMatrix\) obtained as
the output of a sequence of
randomly sampled gates applied to~\(\ket{0}\);
the fidelity is equal to \(\trace{(\rho \dyad{0})}\).

Unitary 2-designs are not the only gate sets that can be used to characterise using a randomised benchmarking scheme.
Using the techniques of randomised benchmarking,
an arbitrary gate set can be characterised~\cite{helsen22prx}. 
However, such characterisation could in principle require up to $d^4$ parameters.
This is part of the importance of dihedral benchmarking for qubits~\cite{dugas2015} and our work for qudits.

Currently,
the only methods to estimate the average gate fidelity of universal gates are established for
qubits:
dihedral benchmarking~\cite{barends2014,dugas2015,Cross2016} and
non-Clifford interleaved benchmarking~\cite{Harper_Flammia_2017};
we summarise these two schemes in this and the next paragraph.
In dihedral benchmarking (DB), single- and multi-qubit gates are benchmarked
through
the average gate fidelity computed over a gate set labelled by elements of the dihedral group~\(D_8\).
The expression for the average gate fidelity and the sequence fidelity in DB depends on
two parameters irrespective of the number of qubits.
A different method to benchmark qubit T gates is achieved by qubit interleaved
benchmarking for the Clifford gate set.
In that scheme, two different circuit designs are used to estimate independently the average gate fidelity of a
T~gate~\cite{Harper_Flammia_2017}.

Interleaved benchmarking uses circuits combining Clifford
and non-Clifford gates. The noise of the composition is characterised. Then, from
an approximation for the fidelity of a composition of channels as a product of
the compositions, the noise of the non-Clifford gate is
obtained~\cite{Harper_Flammia_2017}.

A recent scheme that is more general than randomised benchmarking is shadow
estimation~\cite{Helsen2023}.
For this scheme, real irreducible representations (of a group) are fundamental.
Our work, specifically the representation of the rHDG we construct in
\S\ref{sec:approach},
can also be used---substituting for the Clifford gate set---in such a scheme to characterise universal gates.

Our scheme inherits the advantages of randomised benchmarking schemes.
The reason of this inheritance is because, thanks to our analysis, the scheme is similar to the one
originally done for the Clifford gate set~\cite{magesan2011}.
The advantages of randomised benchmarking are twofold: independence of SPAM
errors~\cite{Helsen_Xue_Vandersypen_Wehner_2019}
and
scalability~\cite{Wright2019}.

We have constructed our scheme around the irreducible representation of a group
we introduce as the real hyperdihedral group (rHDG).
We construct this irrep using two representations.
The first auxiliary representation corresponds to all possible permutations of
the diagonal entries of (a representation of) a T gate (a member of the third
level of the Clifford hierarchy which is not a Clifford operation) and the products that can
be formed from these permutations.
The second representation is generated by two gates---a qubit X gate
(transition between only two levels in a \(\qudim\)-level system) and a qudit X gate.
From these representations we create a faithful irrep of the \thegroup.

Our scheme has the following applications.
Primarily, it benchmarks single- and multi-qudit T gates.
However, 
our scheme can also be used to 
characterise any 
gate of order~\(\QuditDim\) diagonal in
the computational basis
satisfying the following condition:
in terms of the generators of the diagonal matrices, 
each diagonal entry parametrised by the generators of the product of cyclic
groups, is not a linear combination of the other entries.
Additionally, the representation of the rHDG we construct can be exploited in
shadow estimation to characterise a universal gate set without requiring an
inversion gate.
  We emphasise that,
  although our scheme generalises dihedral benchmarking,
  our techniques in this generalisation are distinct and are of interest by themselves.

A few clarifications of the generality of our scheme is needed.
First, only for prime and powers of primes level systems the gates in
\(\cliffordset_3\setminus\cliffordset_2\) are universal.
For the rest of positive integers, Clifford is not a unitary 2-design~\cite{Graydon2021}.
However,
we are convinced that once a universal gate set is introduced for non-powers of
prime systems, our scheme will retain its usefulness to characterise a
non-Clifford diagonal gate.

Whereas using a gate set that is a unitary 2-design (U2D) simplifies
the analysis of a randomised benchmarking scheme,
gate sets that are non-U2D can also be used to characterise
gates~\cite{Kong_2021,helsen22prx}.
For instance, auxiliary schemes such as shadow estimate~\cite{Helsen2023}
and character randomised benchmarking~\cite{helsen22prx}
allow 
the recovery and extension of
most of the basic features of randomised
benchmarking schemes, such as SPAM error independence.

In this paragraph we briefly discuss the differences of our scheme with respect
to Clifford randomised benchmarking.
First, the gate set required is not a unitary 2-design.
This will be shown to imply that our scheme requires two parameters to fit;
the only other experimental change is that two initial states are required.
To characterise a universal gate set, it is still necessary to characterise a
Fourier matrix; this can be done with randomised
benchmarking~\cite{Jafarzadeh_Wu_Sanders_Sanders_2020}.
As we show hereafter,
our scheme is useful to characterise gates,
besides non-Clifford,
corresponding to diagonal unitary matrices.


We now describe the structure of our paper.
Section~2 is background: we introduce the necessary notation,
and explains the key concepts—from representation theory, quantum information,
and randomised benchmarking—essential for the rest of the paper.
Section~3 is devoted to the definition of the real hyperdihedral group (rHDG),
which we use in our scheme to characterise audit T gates.
In Section~4, we compute the expressions for the average gate fidelity and sequence fidelity,
and briefly discuss potential extensions of our research,
highlighting its implications and possibilities for future studies. 

Because our work is long and we expect it to be of interest to both experimental and theoretical groups,
we explain two roadmaps for our work:
one for those readers interested on an experimental implementation and the other for 
those interested on our theoretical results.
The reader more interested on the implementation of the scheme,
\S\ref{subsec:experimental-scheme} and \S\ref{sec:numerics} contain an illustration of the scheme and the explicit resources (in terms of the number of shots,
circuits,
and states) required for the implementation of our scheme.
In that subsection,
we refer to the expressions that are of interest for the characterisation.
For a theoretical audience,
\S\ref{sec:approach} and,
more importantly,
\S\ref{sec:results} are of interest since,
in those sections,
the gate set is discussed both in creation and application in randomised benchmarking schemes.

\section{Background}
\label{sec:background}
In this section we provide the essential background for our benchmarking scheme
for universal qudits gates.
First we set the scene with the basic concepts and accompanying mathematics.
Then we discuss universal gates and channels.
Finally, we go through key concepts for randomised benchmarking schemes and
mathematics of average gate fidelity and sequence fidelity.
\subsection{Setting the scene}
In this subsection
we introduce notation for the mathematical objects needed for our scheme.
Our scheme is developed for~\(\QuditDim\)-dimensional Hilbert spaces~\(\mathsf{H}_\QuditDim \coloneqq\mathrm{span}(\ket{i}\colon i\in\Ring_\QuditDim)\),
with~\(\Ring_\QuditDim\) integers modulo~\(\QuditDim\)~\cite{dummit2018}.
Because currently there is no clearly defined universal gate
set for non-powers of primes, we restrict \(\qudim\) to be a power of prime.

Let~\(\SetOfMatrices_{\QuditDim,\QuditDim}\) denote the set of
\(\QuditDim\)-dimensional square matrices
with complex entries.
Now, let us consider a subset of~\(\SetOfMatrices_{\QuditDim,\QuditDim}\):
we denote by~\(\DensityMatrices'(\QuditDim)
\coloneqq\{\DensityMatrix\in\SetOfMatrices_{\QuditDim,\QuditDim}\colon
\trace{\DensityMatrix} = 1, \DensityMatrix\geq 0, 
\DensityMatrix\text{ is hermitian}
\}\) the set of density matrices for~\(\QuditDim\)-dimensional systems.

We now recall channels and how they act on states.
Consider a finite set of matrices~\(\SetOfKraus \subset \SetOfMatrices_{\QuditDim,\QuditDim}\)
satisfying 
\begin{equation}\label{eq:definition-condition-kraus-operator}
 \sum_{A\in \SetOfKraus} A^{\dagger}A = \mathbb{I}_\QuditDim.
\end{equation}
Then the~\(\QuditDim^2\)-dimensional squared matrix
\begin{equation}\label{eq:kraus}
 \Channel_{\SetOfKraus} \coloneqq \sum_{A\in \SetOfKraus} A\otimes \bar{A}
\end{equation}
(the upper bar denotes complex conjugation) is the channel corresponding to a
so-called set of Kraus operators~\(\SetOfKraus\)
~\cite{kraus1983, HeinosaariTeiko2012Tmlo}.
We define the set
of linear CPTP mappings
\begin{equation}
\SetOfChannels_\QuditDim \coloneqq 
\{\Channel_{\SetOfKraus}\colon
 \SetOfKraus\subset\SetOfMatrices_{\qudim,\qudim},
 \sum_{A\in \SetOfKraus} A^{\dagger} A = \mathbb{I}_\QuditDim
\}.
\end{equation}
When the context is clear,
s
we obviate the operators~\(\SetOfKraus\), 
drop the subindex~\({}_{\SetOfKraus}\),
and only write~\(\Channel\in \SetOfChannels_\QuditDim\).

The action of a channel~\(\Channel\) on a state~\(\DensityMatrix\) is
conveniently computed by first
reshaping~\(\DensityMatrix\) by stacking the columns of~\(\DensityMatrix\),
an operation known as
vectorisation.
We denote the vectorisation of~\(\DensityMatrix\) by~\(\kket{\DensityMatrix}\).
The vectorisation reshaping satisfies, for~\(A\in
\SetOfMatrices_{\QuditDim,\QuditDim}\) and~\(\DensityMatrix \in
\DensityMatrices'_\QuditDim\),
\begin{equation}
\kket{A \DensityMatrix A^{\dagger}} = (A\otimes \bar{A}) \kket{\DensityMatrix}.
\end{equation}
We also denote~\(\DensityMatrices_\QuditDim \coloneqq \{\kket{\DensityMatrix}\colon \DensityMatrix\in
\DensityMatrices'(\QuditDim)\}\)
with~\(\DensityMatrices_\QuditDim\) a Hilbert space with the inner product in~\(\mathbb{C}^{2d}\).
We write the bra of~\(\kket{\DensityMatrix}\) as~\(\bbra{\DensityMatrix}\).

\subsection{Pauli and Clifford gates}\label{subsec:pauli-clif}

Some of the basic operations that can be performed on~\(\mathsf{H}_\QuditDim\), are represented by
the clock and shift
matrices~\({}_\qudim X\) and~\({}_\qudim Z\)~\cite{schwinger, sylvester}.
Let \(k\) be any positive integer,
and \([k] \coloneqq \{0,1,\ldots, k-1\}\).
The matrices~\({}_{\QuditDim}X\) and~\({}_{\QuditDim}Z\) 
act on the computational basis of a~\(\QuditDim\)-dimensional Hilbert space as
\begin{equation}
{}_{\QuditDim} X \ket{i} \coloneqq \ket{i+1}, \qquad
{}_{\QuditDim} Z \ket{i} \coloneqq \RootUnity_{\QuditDim} ^{i} \ket{i},
\end{equation}
with~\(\RootUnity_{\QuditDim} \coloneqq \exp(2\pi \myi/\QuditDim)\),
\(i\in [\qudim]\), 
and addition is defined modulo~\(\qudim\).

From the clock and shift matrices, 
the Heisenberg-Weyl (HW) matrices are defined.
Let~\(i,j\in \Ring_{\QuditDim} \).
Then the HW matrix~\({}_{\QuditDim} W_{di+j}\) is
\begin{equation}\label{eq:hw-matrix}
 {}_{\QuditDim} W_{di+j} \coloneqq {}_{\QuditDim} X^{i}{}_{\QuditDim} Z^{j},
\end{equation}
and the set of HW matrices is denoted by 
\begin{equation}
{}_\QuditDim \mathsf{W}
\coloneqq \{
{}_{\QuditDim} W_{di+j} 
\colon i,j \in \Ring_{\QuditDim} \}.
\end{equation}
In turn,
a unitary generalisation of the Pauli group is~\cite{Patera_Zassenhaus_1988}:
\begin{equation}
 \mathcal{P}_\QuditDim \coloneqq\left\langle {}_\QuditDim W_i\colon i \in\left[\QuditDim^2\right]\right\rangle,
\end{equation}
where we use angular brackets 
to denote the set generated by the argument of the brackets.

Another group highly relevant to our work is the Clifford group.
We denote the subset of~\(\SetOfMatrices_{\QuditDim,\QuditDim}\) formed by
unitary matrices by~\(\mathrm{U}(\QuditDim) \coloneqq \{u\in \SetOfMatrices_{\QuditDim,\QuditDim}\colon u u^{\dagger} = u^{\dagger}u  =
\mathbb{I}_\QuditDim\}\).
The qudit Clifford group~\({}_\QuditDim\mathcal{C}_2\) is the normaliser of~\(\mathcal{P}_\QuditDim\):
\begin{equation}
 {}_\QuditDim\mathcal{C}_2
 \coloneqq
 \{
u\in \mathrm{U}(\QuditDim)\colon
\forall p\in \mathcal{P}_\QuditDim, u p u^{\dagger}\in \mathcal{P}_\QuditDim
 \}.
\end{equation}
We also need the third level of the Clifford hierarchy:
\begin{equation}
 {}_\QuditDim\mathcal{C}_3
 \coloneqq
 \{
u\in \mathrm{U}(\QuditDim)\colon
\forall p\in \mathcal{P}_\QuditDim, u p u^{\dagger}\in {}_\QuditDim\mathcal{C}_2
 \}.
\end{equation}
We omit the subindex~\({}_\QuditDim\) when the dimension of the system is
clear from the context or valid for any~\(\QuditDim\).

We require the multi-qudit version of~\(\mathcal{P}_\QuditDim\), \(\mathcal{C}_2\),
and~\(\mathcal{C}_3\).
Let~\(\numberOfQudits\in \integers\) be the number of qudits.
The multi-qudit Heisenberg-Weyl matrices, which are Clifford, and third-level of the Clifford hierarchy, are 
\begin{subequations}
\begin{align}
 \mathcal{P}_\QuditDim^{ \numberOfQudits} &\coloneqq \{ P_1\otimes \cdots
 \otimes P_\numberOfQudits\colon P_1,\ldots, P_\numberOfQudits\in \mathcal{P}_\QuditDim \},\\
 {}_\QuditDim\mathsf{W}^{\numberOfQudits} &\coloneqq \{ W_1\otimes \cdots
 \otimes W_\numberOfQudits\colon W_1,\ldots, W_\numberOfQudits\in {}_\QuditDim\mathsf{W} \},\\
 {}_\QuditDim\mathcal{C}_2^{\numberOfQudits} &\coloneqq \{ C_1\otimes \cdots
 \otimes C_\numberOfQudits\colon C_1,\ldots, C_\numberOfQudits\in {}_\QuditDim\mathcal{C}_2 \},\\
 {}_\QuditDim\mathcal{C}_3^{\numberOfQudits} &\coloneqq \{ C_1\otimes \cdots
 \otimes C_\numberOfQudits\colon C_1,\ldots, C_\numberOfQudits\in {}_\QuditDim\mathcal{C}_3 \}.
\end{align}
\end{subequations}
Having discussed channels in general, we now focus on important subsets of
channels.
\subsection{Universal gates}
\label{subsec:universal-gates}
Universality is firstly a concept defined for a subset of unitary matrices, which is then
lifted to a set of gates~\cite{Kitaev_1997}.
Let~\(\{V_i \in \mathrm{U}(\QuditDim)\}\) be a finite set of unitary matrices.
Let~\(\mathsf{V} \coloneqq \langle V_i\rangle\) be the group of matrices
generated by the set of matrices~\(\{V_i\}\).
Then~\(\mathsf{V}\) is universal if and only if~\(\operatorname{closure}(\mathsf{V})\cap\mathrm{U}(\qudim) =
\mathrm{U}(\QuditDim)\)~\cite{MikeAndIke,Dawson_Nielsen_2005},
meaning an arbitrary element in $\mathrm{U}(\qudim)$ can be generated to
arbitrary precision by some finite sequence of elements in $\mathsf{V}$.

Each matrix~\(V \in \mathsf{V}\) induces a channel~\(\Channel_{V}
\coloneqq V\otimes \bar{V}\). 
Since $V$ is by construction unitary, we
 refer to the unitary channel~\(\Channel_{V}\) as the 
 representation of the ideal $V$ gate, or as the ideal~\(V\) gate for short.
In the same manner, given a set of unitary matrices, such as the Clifford or Pauli groups,
we can talk about Clifford or Pauli gates.
Therefore, given the clarifications made in the previous two sentences, 
a universal gate set 
refers to the physical implementation of a set of matrices
\(\mathsf{V}\), such that \(\mathrm{closure}(\mathsf{V}) = \mathrm{U}(\QuditDim)\).

There are two single-qudit gates useful for universality purposes.
The qudit Hadamard gate H (more appropriately named Fourier matrix hereafter),
which is a Clifford gate corresponding to the matrix
with entries~$H_{i,j} \coloneqq \RootUnity_{\QuditDim} ^{ij}/\sqrt{\QuditDim}$, is one of two generators of
a universal gate set.
The other generator, which is outside the Clifford group, is~\(T\in \mathcal{C}_3
\setminus \mathcal{C}_2\).

In our work, we only consider T gates diagonal in the computational basis.
For concreteness, we use the following qudit T gates:
\begin{equation}\label{eq:qudit-t-gate}
{}_\qudim T \coloneqq 
\begin{cases}
 \sum_{j\in [\qudim]} \delta_{j,j} \RootUnity_\qudim^{j^{3}},\quad \qudim\neq 2,3,6, \\
\sum_{j\in [\qudim]} \delta_{j,j} \RootUnity_{3\qudim}^{j^{3}},\quad \qudim = 3,6.
\end{cases}
\end{equation}

From the Clifford gate set, only the Hadamard gate is necessary to construct a
universal gate set.
In other words:
\(\mathrm{closure}\langle H,T\rangle = \mathrm{U}(\QuditDim)\)~\cite{Watson2015},
where~\(H\) is the qudit Hadamard matrix and~\(T\in \mathcal{C}_3 \setminus \mathcal{C}_2\). 
For the multi-qudit case,
we need to add, to a universal gate set,
entangling control gates between every pair of qudits to obtain a multi-qudit
universal gate set~\cite{MikeAndIke,Barenco1995}.

The importance of universal gate sets is three-fold~\cite{MikeAndIke}:
first, (as previously mentioned) any unitary matrix can be approximated by the product of a finite set of
matrices from the universal set;
second, every gate acts only on up to three qudits simultaneously~\cite{Reck1994};
third, by the Solovay-Kitaev theorem, 
the sequence of primitive gates necessary to approximate any gate is both efficient to compute on a
classical computer and also efficient in the number of primitive gates~\cite{MikeAndIke,Dawson_Nielsen_2005, Kitaev_1997}.
\subsection{Representation theory detour}\label{subsec:representation-theory-detour}

Representations of a finite group are key to our scheme~\cite{serre1977, Tinkham92}.
Given a finite group~\(\FiniteGroup\),
a~\(\QuditDim\)-dimensional representation of~\(\FiniteGroup\) acting
on~\(\DensityMatrices_{\QuditDim,\QuditDim}\) 
is a map from the abstract group elements to the \(\SetOfMatrices_{\QuditDim,\QuditDim}\) matrices so multiplication of group elements is preserved: 
\begin{align}\label{eq:def-representation}
\mu\colon \FiniteGroup\to \SetOfMatrices_{\QuditDim,\QuditDim}\colon 
\GroupElement &\mapsto \mu(\GroupElement) \nonumber \\
g_1g_2&\mapsto \mu(g_1 g_2)=\mu(g_1)\mu(g_2)\,;
\end{align}
following Serre~\cite{serre1977}, we call~\(\mu(\GroupElement)\) a representative of~\(\GroupElement\) . 
We now recall the notion of reducible and irreducible
representations~\cite{serre1977}.

There may be a change of basis so the matrices $\mu$ of Eq.~\eqref{eq:def-representation} are made block diagonal; the new basis vectors are thus transformed by the block-diagonal $\mu$ into a subset spanning disjoint subspaces.
In such a case, $\mu$ is said to be reducible.
If no change of basis can further block-diagonalize $\mu$,
then each block said to be an irreducible representation of the group.
Note that an irreducible block may occur more than once in the reduction of the original representation.
More formally: 
 assume there is a list of  subspaces~\(\{\Irrep_i\}\) of dimension $\ge 1$ such
that~\(\mu(\GroupElement)\Irrep_i\subset\Irrep_i\),
\emph{v.g.} $\mu(\GroupElement)$ transforms vectors in the block $\Irrep_i$ only into combinations of other vectors in $\Irrep_i$
 and there is no smaller subspace~\(\Irrep_i'\subset
\Irrep_i\) such that~\(\mu(g)\Irrep_i'\subset \Irrep_i'\).
Then we can write 
 \begin{equation}
\DensityMatrices_\QuditDim \cong \bigoplus_i \Irrep_i.
\end{equation}
In a common abuse of language~\cite{serre1977}, the subspace~\(\Irrep_i\) is
also called irreducible representation~\cite{serre1977}.

An example of such a reduction is the 4-dimensional space spanned by two spin-1/2 states,
which can be broken into a one-dimensional subspace and a 3-dimensional subspace
so $U(2)$ transformations only transform triplet states (spin 1 states) amongst themselves,
and transform the singlet (spin 0 states) to a multiple of
itself~\cite{MikeAndIke}.

We define an important representation in our scheme.
For some~\(\QuditDim\)-dimensional 
unitary irreducible representation (unirrep)~\(\UniRepHdg\) of \(\FiniteGroup\), 
we define the representation~\(\Channel_{\UniRepHdg}\) as
\begin{equation}\label{eq:rep_channel_gamma}
\Channel_{\UniRepHdg} \colon 
\FiniteGroup\to \SetOfMatrices_{\QuditDim^2,\QuditDim^2}\colon
\GroupElement\mapsto
\UniRepHdg(\GroupElement)\otimes \overline{\UniRepHdg(\GroupElement)}.
\end{equation}
In our case, $\UniRepHdg(\GroupElement)\otimes \overline{\UniRepHdg(\GroupElement)}$ will be reducible, as shown below.
We also define the noisy representative of~\(\Channel_\UniRepHdg\).
Let~\(\widetilde{\Channel}\in \SetOfChannels_\QuditDim\) be as in~\eqref{eq:kraus}. 
Then the noisy representative of~\(\Channel_{\UniRepHdg}\) is
\begin{equation}\label{def:noise-term}
 \widetilde{\Channel}_{\UniRepHdg}
 \colon 
 \FiniteGroup\to \SetOfMatrices_{\QuditDim^2,\QuditDim^2}
 \colon
 \GroupElement \mapsto
 \widetilde{\Channel}
 \circ
 \Channel_{\UniRepHdg}(\GroupElement);
\end{equation}
the symbol \(\circ\) denotes composition of mappings.
It is notationally economical to
shorten~$\tilde{\Channel}_{\UniRepHdg}(\GroupElement)$ to~$\tilde{\Channel}_g$
in situations where knowledge of~$\UniRepHdg$ is superfluous.
Note that,
because of the noise,
the set $\{\tilde{\Channel}_g\}$ no longer forms a group in the sense that
$\tilde{\Channel}_{g_1}\circ\tilde{\Channel}_{g_2}\ne \tilde{\Channel}_{g_1g_2}$ in general.
Exploiting this feature allow us to recover information about the noise and the average gate fidelity.

\subsection{Average gate fidelity and sequence fidelity}
\label{subsec:agf-sf-expressions-fidelity-sequence-gate}
Consider a set  of channels labelled by the elements of a finite
group~\(\FiniteGroup\).
Standard RB schemes estimate
the AGF (or related quantities such as process fidelity and entanglement fidelity)
over~\(\FiniteGroup\)~\cite{Proctor2017}.
We now recall the formula for the~AGF used in gate-independent RB schemes.

\begin{lemma}[\cite{NIELSEN2002249,Horodecki90}]
Let~\(\FiniteGroup\) be a finite group with a~\(\QuditDim\)-dimensional
representation~\(\UniRepHdg\).
Let~\(\tilde{\Channel}_g\) be the noisy representative of~\(\GroupElement\).
Then the average gate fidelity of~\(\tilde{\Channel}_\GroupElement\) (with
respect to~\(\Channel_\GroupElement\)) is computed as
\begin{equation}
 \Fidelity\colon
 (\Channel_\GroupElement, \tilde{\Channel}_\GroupElement) \mapsto
 \frac{1}{\QuditDim(\QuditDim+1)} \trace(\tilde{\Channel}_\GroupElement^{\dagger}\Channel_\GroupElement)
 +
 \frac{1}{\QuditDim+1}.
\end{equation}
\end{lemma}

We now discuss the quantity randomised benchmarking schemes allow to estimate.
 Let~\(\FiniteGroup\) be a finite group that labels ideal and noisy
 channels---\(\Channel_\UniRepHdg\) and~\(\tilde{\Channel}_\UniRepHdg\).
Then the average gate fidelity over~\(\FiniteGroup\) is
\begin{equation}
 \Fidelity(\FiniteGroup)
 \coloneqq 
 |\FiniteGroup|^{-1}
 \sum_{\GroupElement\in \FiniteGroup} 
\Fidelity(\Channel_\GroupElement, \tilde{\Channel}_\GroupElement).
\end{equation}
For clarity in our exposition
we present our results for the case of gate-independent noise, but 
we note that our results can be extended to the case where every gate has
different noise~\cite{Wallman_2018, Merkel_Pritchett_Fong_2021}.

\section{Approach}
\label{sec:approach}
We now introduce the key ingredient of our generalisation of single- and
multi-qudit dihedral benchmarking: the qudit real hyperdihedral group (\thegroup).
First we introduce auxiliary groups needed for the definition of the~\(\thegroup\). 
Then we construct representations for the auxiliary groups essential to the
definition of a unirrep for the~\(\thegroup\);
from this unirrep, our expressions for the gate fidelity and sequence fidelity are derived.
This is a good place to make the important observation that
the \thegroup's construction is valid for any diagonal gate.
Thus, our scheme is also useful to characterise any diagonal
gate with order equal to, or greater than 
two, such that its parametrisation does not have linearly independent entries.
Alternatively,
any diagonal matrix with order equal to, or greater than 
\(\Order(\qudim)\)
as defined in Eq.~\eqref{eq:order-t-gate}
can also be characterised. 
We conclude this section by presenting the decomposition of the 
qudit Hilbert space in irreps of~\(\thegroup\).

\subsection{Constructing the rHDG}\label{subsec:construction-rhdg}
The~\(\thegroup\) is defined in terms of
the symmetric group of~\(\QuditDim\) elements and 
cyclic groups~\cite{Tinkham92}.
We denote by~\(S_{\QuditDim} \) the group of permutations of~\(\QuditDim\) elements
and~\(C_k\) the cyclic group of order~\(k\).

The \thegroup\ requires a product of cyclic groups.
The number of cyclic groups in the product and their order depends
on~\(\QuditDim\) and~\({}_\qudim T\), with~\({}_\qudim T\) given in Eq.~\eqref{eq:qudit-t-gate}.
We proceed to describe the structure of an arbitrary finite product of cyclic
groups.
After introducing the notation, we explain how to construct the specific product
for a given diagonal gate~\(T\).

Some preliminary definitions are necessary before our discussion of the product
of cyclic groups.
Let~\(\Order(\qudim)\)
denote the order of~\({}_\qudim T\) in Eq.~\eqref{eq:qudit-t-gate}; that is,
\begin{equation}\label{eq:order-t-gate}
 \Order: 
 \QuditDim
 \mapsto
\begin{cases}
 9 & \QuditDim = 3, \\
 18 & \QuditDim = 6, \\
 \QuditDim & \QuditDim\neq 3,6
\end{cases}.
\end{equation}
Then for some~\(l\in \integers\) and~\(\bm{k} = (k_\startnumbering, \ldots, k_l)\in \Ring_{\Order(\QuditDim)}^{l}\),
we write the product of cyclic groups as
\begin{equation}\label{eq:structure-product-cyclic-groups}
\bigtimes_{k\in \bm{k}} C_k
\coloneqq 
C_{k_\startnumbering} \times \cdots
\times
C_{k_l}.
\end{equation}
The structure in Eq.~\eqref{eq:structure-product-cyclic-groups} depends on the
form of \(T\); we now discuss how to compute this product of cyclic groups.

For a given~\(\QuditDim\) and~\(T\),
we compute the necessary product~\( \bigtimes_{k\in \bm{k}} C_k \)
using Howell's algorithm.
This algorithm
is a generalisation of the row-echelon reduction algorithm to matrices
with entries in an arbitrary finite ring.
An example of the computation of~\( \bigtimes_{k\in \bm{k}} C_k \)
for qutrits is provided in Appendix~\ref{app:product-cyclic-groups}.

We continue our construction of \thegroup\ by introducing representations for
the auxiliary groups.
First we define the \(\qudim\times\qudim\) representation for the symmetric group
\begin{equation}\label{eq:def-representation-symmetricgroup}
 \RepSymm\colon S_{\QuditDim} \to
 \SetOfMatrices_{\QuditDim,\QuditDim}\colon \Permutation \mapsto
 \sum_{i=1}^{\QuditDim} \delta_{\Permutation(i), i}.
\end{equation}
These \(\qudim!\) matrices have only one non-zero entry in each row or column,
and the non-zero entry is 1;
the representation \(\RepSymm\) is known in the mathematical literature as the
standard representation, or the permutation representation~\cite{serre1977}.

To complete the definition of the representation of the second auxiliary group,
we introduce two sets related to a matrix~\(T\).
Let
\begin{equation}\label{eq:def-permutations-t-gate}
  \bm{T}' \coloneqq \langle \RepSymm(\Permutation) T
  {\RepSymm(\Permutation)}^{\dagger}\colon \Permutation\in S_{\QuditDim} \rangle
\end{equation}
be the set resulting from permuting the diagonal entries of the qudit~\(T\).
By the procedure detailed in Appendix~\ref{app:product-cyclic-groups},
we extract from~\(\bm{T}'\) the minimal generating set 
\begin{equation}\label{eq:isomorphism-product-cyclic}
\bm{T} \coloneqq (T_\startnumbering,\ldots, T_l),
\end{equation}
where each~\(T_i\) has order at most~\(\Order(\QuditDim)\).

Using~\(\bm{T}\), we introduce the second auxiliary representation.
Let us consider
an element~\(\bm{\CyclicElement} = (\CyclicElement_\startnumbering, \ldots,
\CyclicElement_l)\in
\bigtimes_{k\in \bm{k}}C_{k}
\).
The representation of~\(\bm{T}'\) is 
\begin{equation}\label{eq:def-representation-product-cyclic-groups}
\RepCyclic\colon \bigtimes_{k\in \bm{k}} C_{k} \to
\SetOfMatrices_{\QuditDim,\QuditDim}\colon
\bm{\CyclicElement} \mapsto \prod_i T_i^{\CyclicElement_i},
\end{equation}
where---in an abuse of notation---\(\CyclicElement_k\) can also denote a member of the
ring~\(\Ring_{k}\)~\cite{dummit2018}.

We are now ready to define the \thegroup\ and its unirrep that underpins our
scheme to characterise qudit T gates.
The \thegroup\ is 
\begin{equation}\label{eq:the-group}
\thegroup \coloneqq S_{\QuditDim} \ltimes \bigtimes_{k\in \bm{k}} C_k;
\end{equation}
where the symbol \(\ltimes\) denotes 
semidirect product~\cite{dummit2018,GAP4}.
We use the semidirect product to formally consider the
multiplication rule between elements of the \thegroup\ is 
\begin{equation}
\RepSymm(\sigma_1)\RepCyclic(\alpha_1)
\RepSymm(\sigma_2)\RepCyclic(\alpha_2)
=
\RepSymm(\sigma_1\sigma_2)
\RepCyclic(\alpha_1^{\sigma_2})
\RepCyclic(\alpha_2),
\end{equation}
noting that 
\(
\RepCyclic(\alpha_1^{\sigma_2})
\coloneqq 
\RepSymm(\sigma_2)^{\dagger}\RepCyclic(\alpha_1)\RepSymm(\sigma_2)\)
is also a diagonal matrix
and is a representative of an element in \(\bigtimes_{k\in \bm{k}} C_k\).
In turn, 
the unirrep~\(\UniRepHdg\) for \thegroup\ is
\begin{equation}\label{eq:representation-gamma-unirrep-rhdg}
\UniRepHdg \colon
\thegroup \to \SetOfMatrices_{\QuditDim,\QuditDim}
\colon
(\Permutation,\bm{\CyclicElement})
\mapsto
\RepSymm(\Permutation) \RepCyclic(\bm{\CyclicElement}).
\end{equation}
Next we describe the decomposition of the Hilbert space~\(\DensityMatrices_\QuditDim\) in terms of irreps of~\thegroup.

We now discuss a fundamental point: 
the decomposition of 
the Hilbert space \(\DensityMatrices_\QuditDim\) in terms of irreps of \thegroup.
The representation~\(\Channel_\UniRepHdg = \UniRepHdg\otimes\bar{\UniRepHdg}\)
defined in Eq.~\eqref{eq:rep_channel_gamma}
decomposes the Hilbert space~\(\DensityMatrices_\QuditDim\) into three irreps;
we prove this fact
in
Appendix~\ref{appendix:properties-rhdg}.
These representations are:~\(\Irrep_{\mathbb{I}}\), the trivial irrep;
\(\Irrep_0\), the standard irrep (of dimension \(\qudim-1\));
and \(\Irrep_+\), the complement of~\(\Irrep_{\mathbb{I}}\) and~\(\Irrep_0\).
We write the decomposition as
\begin{equation}\label{eq:irrep-decomposition}
\DensityMatrices_\QuditDim 
\cong
\Irrep_{\mathbb{I}}
\oplus
\Irrep_0
\oplus
\Irrep_+.
\end{equation}
In the following section we use the decomposition in
Eq.~\eqref{eq:irrep-decomposition} to define our RB scheme.

The fact that the representations~\(\Irrep_{\mathbb{I}}, \Irrep_0, \Irrep_+\)
are real makes our gate set
\thegroup\
useful for other schemes besides randomised benchmarking. 
The scheme known as
shadow estimation~\cite{Helsen2023} can be used to characterise a gate set.
The only requirement for a gate set, in the shadow estimation scheme, 
is that the Kraus-channel representation decomposes into real irreps.
Therefore, using our rHDG gate set, shadow estimation can be used to
characterise a qudit universal gate set.

Our scheme also serves to characterise hybrid platforms~\cite{Daboul2003}.
A hybrid platform is a combination of two qudits:
one is a \(d_1\)-level system 
and the other a \(d_2\)-level system, with \(d_1\neq d_2\)
and also primes or powers of primes.
By considering the X and T gates for each subsystem, 
one of dimension \(d_1\) and the other of dimension \(d_2\),
we construct the \thegroup\ 
for the combined system. The Kraus-representation of these gates decomposes as in
Eq.~\eqref{eq:irrep-decomposition}.  Thus, it is possible to use our scheme to
estimate the average gate fidelity of hybrid platforms.

In this section we introduced one irrep for the rHDG that
leads, by using the associated Pauli-Liouville representation,
to the decomposition of the Hilbert
space~\(\DensityMatrices_\QuditDim\) into three irreps.
Our results are valid not only for a qudit T gate but for any diagonal unitary
matrix with order at least~2,
such that the parametrisation of a representative of the product of cyclic
groups has no linearly independent entries.
In the following section we use the decomposition in Eq.~\eqref{eq:irrep-decomposition}
to define our RB scheme.
We principally use the projectors onto the irreps in Eq.~\eqref{eq:irrep-decomposition}.

\section{Results}
\label{sec:results}
In this section, we describe the algorithm used to estimate the AGF of a rHDG gate set.
By using a rHDG gate set and the standard randomised
benchmarking~\cite{magesan2011},
we extend randomised benchmarking to characterise universal gates.
In practice, our scheme is useful to characterise single-qudit T gates and also controlled T
gates;
our scheme also characterises any diagonal unitary gate
with order
greater or equal than~\(\qudim\).

\subsection{Randomised benchmarking and the real hyperdihedral group}
\label{subsec:rb-rhdg}

We start the exposition of the expressions of our scheme by introducing notation.
We denote by~\(\Projector_\IrrepIndex\) the projector onto 
the subspace~\(\Irrep_\IrrepIndex\) of Eq.~\eqref{eq:irrep-decomposition}, 
with~\(\IrrepIndex~\in~\{\mathbb{I}, 0, +\}\).
The projectors~\(\Projector_\IrrepIndex\) appear in our expression for the
\thegroup-twirl of a channel.
\begin{lemma}
Let~\(\Channel\in \SetOfChannels_\QuditDim\) be a qudit channel.
Then
the \thegroup\ twirl of~\(\Channel\) is
\begin{equation}\label{eq:decomposition-rhdg}
 \TwirlOp_{\Channel}
 \coloneqq 
\Eigenvalue_{\mathbb{I}}(\Channel) \Projector_{\mathbb{I}} \oplus
\Eigenvalue_{\mathrm{0}}(\Channel) \Projector_{\mathrm{0}} \oplus
\Eigenvalue_{\mathrm{+}}(\Channel) \Projector_{\mathrm{+}},
\end{equation}
where
\begin{equation}\label{eq:lambda-definition}
\Eigenvalue_\IrrepIndex(\Channel) \coloneqq \frac{\trace (\Channel
\Projector_\IrrepIndex)}{\dim\IrrepIndex},\qquad
\IrrepIndex\in \{\mathbb{I},0,+\}.
\end{equation}
\end{lemma}
\begin{proof}

The expression of the twirl~\(\TwirlOp_{\Channel}\) 
in Eq.~\eqref{eq:decomposition-rhdg}
and the quantities~\(\Eigenvalue_i\) are
consequences of both Schur's lemma~\cite{serre1977} and our decomposition in
Eq.~\eqref{eq:irrep-decomposition}.
The projectors are explicitly constructed in Corollary~\ref{corollary:access-to-parameters}.
\end{proof}

The next two corollaries are our expressions for the sequence and
average gate fidelity.
Notice that specialising our expressions for~\(\qudim=2\),
we recover the fidelities in dihedral~benchmarking for single and multiqubit gates~\cite{dugas2015, Cross2016}.
\begin{corollary}[Expression for the average gate fidelity]
 Let~\(\{\tilde{\Channel}_g\colon \GroupElement\in \thegroup\}\) be a gate set with
 the same noise map~\(\Channel\);
 that is, the noise in Eq.~\ref{def:noise-term} is the same for every group
 element.
 Then the average gate fidelity over~\(\{\tilde{\Channel}_g\colon \GroupElement\in \thegroup\}\)
 is
 \begin{equation}
 \Fidelity(\thegroup) = 
 \frac{\QuditDim 
 (
 \Eigenvalue_{\mathbb{I}} +
 (\QuditDim-1)\Eigenvalue_{0} +
 (\QuditDim^2-\QuditDim) \Eigenvalue_+
 ) + \QuditDim^2}{\QuditDim^2(\QuditDim+1)}.
 \end{equation}
\end{corollary}
\begin{corollary}[Expression for the sequence fidelity]
 Consider the state~\(\ket{\somestate}\),
 the measurement~\(\dyad{\somestate}\),
 and the set of gates---with fixed noise~\(\Channel\)---\(\bm{\channel} \coloneqq \{\tilde{\Channel}_g\colon
 \GroupElement\in \thegroup\}\).
 Let~\(\CircuitDepth\in \integers\) be the circuit depth.
 Then the sequence fidelity for the \thegroup\ is 
\begin{equation}\label{eq:survival-probability}
 \Pr(\CircuitDepth; \ket{\somestate},\dyad{\somestate}, \bm{\channel})
 \coloneqq 
 \bbra{\somestate}
 \Channel
 \TwirlOp_{\Channel}^{\CircuitDepth}
 \kket{\somestate}
 = \sum_{\IrrepIndex\in\{\identity, 0, +\}} 
 \bbra{\somestate}\Channel \Projector_\IrrepIndex
 \kket{\somestate}\Eigenvalue_\IrrepIndex^{\CircuitDepth}.
 \end{equation}
\end{corollary}

A priori, even given the form of the twirl, it is unknown what initial states
and measurements are required for a proper characterisation.
Thus, our next step is to specify the states that our scheme requires.
Let H be the~\(\QuditDim\)-dimensional Hadamard gate and~\(H\) its matrix
representation more appropriately known as Fourier transform matrix.
We define the state
\begin{equation}\label{eq:definition-state-plus}
 \ket{+} \coloneqq F\ket{0}.
\end{equation}
We further simplify the expression for the sequence fidelity in
Eq.~\eqref{eq:survival-probability}.
Let~\(\IrrepIndex\in \{0,+\}\) and let~\(\bm{\channel}\) be a gate set with
fixed noise~\(\Channel\)---the same noise for every group element---such
that~\(\Channel\) is trace-preserving.
The following corollary is proven at the end of Appendix~\ref{appendix:properties-rhdg}.
\begin{restatable}{corollary}{grotten}\label{corollary:access-to-parameters}
 Let~\(\Projector_+\) denote the projector onto the subspace~\(\Irrep_+\).
 Then the projector \(\Projector_+\) maps the vectorisation of  \(\dyad{0}\) to
 the null-vector of \(\DensityMatrices_\qudim\); that is,
 \begin{equation}
 \Projector_+ \kket{0}= \NullVector.
 \end{equation}
\end{restatable}
Then, using Corollary~\ref{corollary:access-to-parameters}, we have the following simplification for the sequence fidelity of Eq.~\eqref{eq:survival-probability}, namely,
\begin{equation}\label{eq:survival-probability-initialstate}
 \Pr(\CircuitDepth; \ket{\IrrepIndex},\dyad{\IrrepIndex}, \bm{\Channel}) =
 A_{\IrrepIndex} + B_\IrrepIndex \Eigenvalue_\IrrepIndex^{\CircuitDepth}, 
\end{equation}
with \(\IrrepIndex\in \{0,+\}\),\(A_\IrrepIndex \coloneqq \bbra{\IrrepIndex}\Channel\Projector_{\mathbb{I}}
\kket{\IrrepIndex} \),~\(B_\IrrepIndex
\coloneqq \bbra{\IrrepIndex}\Channel\Projector_\IrrepIndex \kket{\IrrepIndex} \),
and~\(\Eigenvalue_\IrrepIndex\) given in Eq.~\eqref{eq:lambda-definition}.
Note that the parameters \(\Eigenvalue_\IrrepIndex\)
are `parts' of the trace of the Pauli-Liouville representation of the noise.
Therefore, by estimating these parameters we can estimate the trace of that
Pauli-Liouville representation of the noise;
this is a standard technique in randomised benchmarking
schemes~\cite{MagesanEaswar2012Emoq}.
\subsection{Experimental scheme}\label{subsec:experimental-scheme}
In this subsection,
we describe the experimental arrangement required for the implementation of our
scheme.
Our description is at the level of primitive gates,
state preparation,
and measurement.

For definiteness, we consider the qutrit case. The primitive gates are
\begin{equation}
\RepSymm((23)) = 
\begin{bmatrix}
  1 & 0 & 0 \\
  0 & 0 & 1 \\
  0 & 1 & 0 
\end{bmatrix},
\RepSymm((12)) =
\begin{bmatrix}
  0 & 1 & 0 \\
  1 & 0 & 0 \\
  0 & 0 & 1 
\end{bmatrix},
\text{ and }
T =
\begin{bmatrix}
  \RootUnity_9 &   &   \\
               & 1 &   \\
               &   & 1 
\end{bmatrix};
\end{equation}
the notation \((ij)\) refers to a permutation between the \(i\)-th and \(j\)-th
elements in a list.
We label the elements 
of the \thegroup\ 
by a pair formed by a permutation \(\sigma\) and three values \(\bm{\alpha} = (\alpha_0, \alpha_1, \alpha_2)\) with each \(\alpha_i \in [9]\). 
The following matrices form the gate set:
\begin{equation}
\gamma(\sigma, \bm{\alpha}) \coloneqq \RepSymm(\sigma)
\begin{bmatrix}
  \RootUnity_9^{\alpha_0} & & \\
               & \RootUnity_9^{\alpha_1} & \\
               & & \RootUnity_9^{\alpha_2}
\end{bmatrix},
\end{equation}
where each permutation \(\sigma\) must be written as products of \((12)\) and
\((23)\)
so as to only use the primitive gates.

We now illustrate an example run, taking circuit depth \(m = 3\).
The initial states required to prepare are
\begin{equation}
  \ket{0} \coloneqq
\begin{bmatrix}
1 \\
0 \\
0 \\
\end{bmatrix}
\text{ and }
  \ket{+} = F \ket{0} \coloneqq
  3^{-1/2}
  \begin{bmatrix}
  1 \\
  1 \\
  1
  \end{bmatrix};
\end{equation}
we consider the gate \(F\) in our circuit design.
Then assume the following permutation \(\sigma\) and powers of \(C_9\) are sampled:
\begin{equation}
  \gamma((23), (7,8,8)),
  \gamma((123), (0,5,7)),
  \gamma((12), (8,1,4)).
\end{equation}
To compute the inversion gate, first we determine the inversion permutation:
\begin{equation}
  {(23)(123)(12)}^{-1} = e
\end{equation}
(\(e\) is the identity permutation)
and then the inversion phase gate
\begin{equation}
  \left(
\begin{bmatrix}
  \RootUnity_9^{7} & & \\
               & \RootUnity_9^{8} & \\
               & & \RootUnity_9^{8}
\end{bmatrix}
\begin{bmatrix}
  \RootUnity_9^{0} & & \\
               & \RootUnity_9^{5} & \\
               & & \RootUnity_9^{7}
\end{bmatrix}
\begin{bmatrix}
  \RootUnity_9^{8} & & \\
               & \RootUnity_9^{1} & \\
               & & \RootUnity_9^{4}
\end{bmatrix}
\right)^{-1}
  =
\begin{bmatrix}
  \RootUnity_9^{3} & & \\
               & \RootUnity_9^{4} & \\
               & & \RootUnity_9^{8}
\end{bmatrix};
\end{equation}
then the inversion gate is
\begin{equation}
\begin{bmatrix}
  \RootUnity_9^{3} & & \\
               & \RootUnity_9^{4} & \\
               & & \RootUnity_9^{8}
\end{bmatrix},
\end{equation}
corresponding to the \thegroup\ element \(\gamma(e, (3,4,8))\).
Therefore, a run of the experiment with the previous sequence of gates corresponds to the following circuits:
\begin{figure}[H]
\centering
\begin{equation}\label{eq:circuit_0}
\begin{quantikz}[column sep=5pt, row sep={20pt,between origins}]
  \lstick{\(\ket{0}\)}&
  \gate{\scriptstyle\gamma((23), (7,8,8))}&
  \gate{\scriptstyle\gamma((123), (0,5,7))}&
  \gate{\scriptstyle\gamma((12), (8,1,4))}&
  \gate{\scriptstyle\gamma(e, (3,4,8))}&
\rstick{\(\bra{0}\)}
\end{quantikz}
\end{equation}
\end{figure}
and
\begin{figure}[H]
\centering
\begin{equation}\label{eq:circuit_1}
\begin{quantikz}[column sep=5pt, row sep={20pt,between origins}]
  \lstick{\(\ket{0}\)}&
  \gate{\scriptstyle F}&
  \gate{\scriptstyle \gamma((23), (7,8,8))}&
  \gate{\scriptstyle \gamma((123), (0,5,7))}&
  \gate{\scriptstyle \gamma((12), (8,1,4))}&
  \gate{\scriptstyle \gamma(e, (3,4,8))}&
  \gate{\scriptstyle F^{\dagger}}&
  \rstick{\(\bra{0}\)}
\end{quantikz}.
\end{equation}
\end{figure}

The number of shots refers to the number of times the previous circuits need to be implemented and measured to estimate the quantities
\begin{equation}
\tr{\dyad{0} Q(\dyad{0})}
\text{ and }
\tr{\dyad{+} Q(\dyad{+})},
\end{equation}
where
\begin{equation}
Q =
\tilde{\gamma}((23), (7,8,8))
\tilde{\gamma}((123), (0,5,7))
\tilde{\gamma}((12), (8,1,4))
\tilde{\gamma}(e, (3,4,8))
\end{equation}
are the gates in Eqs.~\eqref{eq:circuit_0} and \eqref{eq:circuit_1}.
After the experimental procedure has been carried out,
the resulting graph circuit~depth \emph{vs} averaged sequence fidelity
is fitted to the curve of Eq.~\eqref{eq:survival-probability-initialstate}.
We leave for \S\ref{sec:numerics} numerical results of our scheme and the statistical analysis of the number of shots and circuits required.

\subsection{Characterisation of multi-qudit gates}\label{subsec:multi-qudit}

In this subsection,
we describe our scheme to characterise multiqudit gates.
Whereas there are currently schemes to characterise Clifford gates in multilevel gates,
there is a lack of methods for non-Clifford gates.
Currently,
there are at least three platforms
with access to entangling gates~\cite{Goss2022,Chi2022,Hrmo2023,Ringbauer2022}.
Since the experimental scheme is the same as for the single-qudit case, 
we do not repeat the discussion.

For bi-qudits, 
the generating set comprises to a single X gate and the CSUM
gate~\cite{Mato2023,Goss2022}, which is a generalised version of the CNOT gate for qudits. This gate is defined as
\begin{equation}
  \mathrm{CSUM}\ket{a,b} = \ket{a,a\oplus b};
\end{equation}
addition is taken modulo \(\qudim\).
This generating set has the same form (and same irrep decomposition) as the rHDG and is therefore the rHDG for bi-qudit systems.

For more than one qudit,
in general,
the gate set requires CSUM gates for each pair of qudits.
The \(n\)-qudit \thegroup\ is
 \begin{equation}
{}_{\qudim^{n}}
\mathrm{rHDG}
\coloneqq 
\langle 
\mathrm{CSUM}_{i,i+1},
X^{(i)}, X^{(\qudim)},
T^{(1)}
\colon i \in [\qudim-1]\rangle,
\end{equation}
where 
\begin{equation}
X^{(i)}
\coloneqq 
\underbrace{\identity\times\cdots\times\identity}_{i-1 \text{ times}}\times X\times
\identity\times\cdots\times\identity
\end{equation}
and 
\(\mathrm{CSUM}_{i,i+1}\)
over every
 \(i \in [\qudim-1]\)
means that we use as control the \(i\)-th qudit and target the \(i+1\)-th qudit.

To conclude this subsection, we briefly comment on a further generalisation.
When the noise is either non-trace preserving or varies across gates, it becomes necessary
to employ the Fourier-transform formalism~\cite{Merkel_Pritchett_Fong_2021, Wallman_2018}.
This formalism is compatible with our decomposition.

\section{Numerics and construction of group elements for qutrits and ququarts}
\label{sec:numerics}
In this section we discuss the number of shots and circuits required
for the estimate of the average gate fidelity for the qutrit and ququart cases.
These results are to be used to determine the feasibility of our scheme.
Our investigations are numerical because the 
variance of the sequence fidelity for the \thegroup\ is currently only possible to compute symbolically (using \texttt{Mathematica}) and not
analytically~\cite{Wallman_Flammia_2014}.

The structure of this section is as follows.
First we discuss the numerical simulations and the interpretation of their outcomes.
Then we discuss the conclusions drawn from the resulting confidence
intervals.
We conclude the section with a summary proposing an experimental strategy 
supported
by our results.
Our simulations approximate the confidence intervals of the 
parameter \(\Eigenvalue_0\)
estimated from a randomised benchmarking experiment using as gate set the
\thegroup.\@

In a randomised benchmarking scheme there are two important quantities
in the design of the experiment:
the number of shots and the nubmer of circuits used.
In this section, we estimate the optimal values for these experimental
parameters based on numerical
simulations.
First, we describe our simulation linking it with an hypothetical experiment.

Given a circuit depth,
we must choose a sequence of gates,
referred to as a circuit.
The value of the fidelity for the initial state under the action of the circuit is called the sequence fidelity.
Thus,
for a given circuit depth \(m\),
we have a list of possible sequences of gates;
if the gate set has order \(|\FiniteGroup|\),
the number of possible sequences is \(|\FiniteGroup|^{m}\).

In an experiment,
only a sequence of \(0\)s and \(1\)s is recorded,
depending on whether the initial state is recovered or not: click or no click.
The distribution of clicks or no clicks follows a Bernoulli distribution.
Thus, for a given circuit, the sequence fidelity is the mean of a list of
Bernoulli distribution. By the central limit theorem,
we can approximate the mean of a list of Bernoulli distributions by a normal
distribution.
In turn, the distribution of the mean sequence fidelity, \(\Pr(\CircuitDepth;
    \ket{\somestate},\dyad{\somestate}, \bm{\channel})\) in
Eq.~\eqref{eq:survival-probability-initialstate},
is the mean of a list of normal distributions. Therefore, \(\Pr(\CircuitDepth;
    \ket{\somestate},\dyad{\somestate}, \bm{\channel})\) is also a normal
distribution, with the mean given by Eq.~\eqref{eq:survival-probability-initialstate}.

Therefore, our numerical simulation can be summarised in the following steps
\begin{enumerate}
  \item Fix circuit depth \(m\), number of shots \(s\), and number of random
    circuits \(r\).
  \item Obtain the sequence fidelity corresponding to the twirl to the power
    \(m\); as argued above, this is the mean of a normal distribution that we
    sample in a future step.
  \item Obtain the variance of the sequence
    fidelity~\cite{Helsen_Wallman_Flammia_Wehner_2019}.
  \item Randomly sample the normal distribution to obtain an approximate
    sequence fidelity \(\check{p}\).
  \item Compute a Bernoulli random variable parametrised by \(\check{p}\).
    Use this Bernoulli random variable to sample a binary sequence of length \(s\).
    The mean is the
    approximate sequence fidelity of an experiment: call it \(p'\).
  \item Repeat the previous steps (ii-v) to average \(r\) different values of
    \(p'\) that approximates~\(\Pr(\CircuitDepth;
    \ket{\somestate},\dyad{\somestate}, \bm{\channel})\).
\end{enumerate}

Concluding the previous steps, we obtain a list of pairs. Each pair is formed by
a circuit
depth and the approximated sequence fidelity.
For this list of values we construct a graph from which we fit using
Eq.~\eqref{eq:survival-probability-initialstate}, obtaining a parameter
\(\check{\eta}_0\).
For convenience, we only study the case for initial state \(\ket{0}\).

Fitting 
(according to Eq.~\eqref{eq:survival-probability-initialstate}) 
the graph constructed as explained in the previous
paragraph
yields an estimate of the parameter \(\eta_0\);
we denote the approximation \(\tilde{\eta}\).
We repeat the simulation step \(10^{3}\) times,
then compute \(\abs{\eta-\tilde{\eta}}\).
From these values we compute quantiles to determine an error and confidence.
The quantiles we use, for the difference \(\abs{\eta_0-\tilde{\eta}_0}\), 
are 0.95, 0.99, and 1.
Then, by reporting the number of shots and the number of randomly drawn
circuits,
paired with the error and confidence, we can provide estimates of the experimental
statistical elements (shots and circuits) necessary for a sensible
characterisation.

We limited our simulations to 10 and 100 cycles 
so as to be a representative figure for an actual experiment;
these numbers are
also within the range of numbers reported in the
literature~\cite{Blok_Ramasesh_Schuster_OBrien_Kreikebaum_Dahlen_Morvan_Yoshida_Yao_Siddiqi_2021}.
In order to assess the spectrum of quality of gates,
our simulation is parametrised by the average gate fidelity of the noise
channel.
We use a random channel~\cite{Kukulski2021}
which is non-unital but CPTP.
The results are presented as follows.

\begin{table}[H]
  \centering
\begin{tabular}{llllll}
Shots & Circuits & \multicolumn{3}{l}{Error given a probability} & Fidelity              \\
\(s\)      & \(r\)          & 0.95          & 0.999         & 1             & \multirow{4}{*}{0.89} \\
      \hline
100   & 10       & 0.013         & 0.020         & 0.024         &                       \\
10    & 100      & 0.013         & 0.021         & 0.021         &                       \\
100   & 100      & 0.003         & 0.006         & 0.008         &                      
\end{tabular}
\caption{\label{tab:tab_1}
Table reporting the confidence values for the estimate of the parameter
\(\eta_0\). The noise 
corresponds to a randomly sampled channel with average gate fidelity \(0.89\).
}
\end{table}

\begin{table}[H]
  \centering
\begin{tabular}{llllll}
Shots & Circuits & \multicolumn{3}{l}{Error given a probability} & Fidelity              \\
 \(s\)     & \(r\)          & 0.95          & 0.999         & 1             & \multirow{5}{*}{0.931339} \\
      \hline
100   & 10       & 0.008         & 0.013         & 0.014         &                       \\
10    & 100      & 0.008         & 0.014         & 0.015         &                       \\
100   & 100      & 0.002         & 0.004         & 0.004         &                       \\
20    & 100      & 0.005         & 0.008         & 0.009         & \\
20    & 20       & 0.013         & 0.022         & 0.022         & 
\end{tabular}
\caption{\label{tab:tab_2}
Table reporting the confidence values for the estimate of the parameter
\(\eta_0\). The noise model is a
composition of a totally depolarising and an amplitude damping channel, which is a comprehensive Markovian and completely positive
noise model. The value of fidelity corresponds to the parameters \(0.01\) for the
depolarising and amplitude damping channel.
}
\end{table}

\begin{table}[H]
  \centering
\begin{tabular}{llllll}
Shots & Circuits & \multicolumn{3}{l}{Error given a probability} & Fidelity              \\
 \(s\)     &  \(r\)         & 0.95          & 0.999         & 1             & \multirow{5}{*}{0.958284} \\
      \hline
100   & 10       & 0.005         & 0.007         & 0.008         &                       \\
10    & 100      & 0.005         & 0.008         & 0.009         &                       \\
100   & 100      & 0.001         & 0.002         & 0.002         &                       \\
20    & 100      & 0.003         & 0.004         & 0.005         & \\
20    & 20       & 0.008         & 0.011         & 0.012         & 
\end{tabular}
\caption{\label{tab:tab_3}
Table reporting the confidence values for the estimate of the parameter
\(\eta_0\). The noise model is a
composition of a totally depolarising and an amplitude damping channel, which is a comprehensive Markovian and completely positive
noise model. The value of fidelity corresponds to the parameters \(0.03\) for the
depolarising and amplitude damping channel.
}
\end{table}

\begin{table}[H]
  \centering
\begin{tabular}{llllll}
Shots & Circuits & \multicolumn{3}{l}{Error given a probability} & Fidelity              \\
 \(s\)     & \(r\)         & 0.95          & 0.999         & 1             & \multirow{5}{*}{0.985921} \\
      \hline
100   & 10       & 0.002         & 0.002         & 0.003         &                       \\
10    & 100      & 0.002         & 0.003         & 0.004         &                       \\
100   & 100      & 0.000         & 0.000         & 0.001         &                       \\
20    & 100      & 0.001         & 0.002         & 0.002         & \\
20    & 20       & 0.003         & 0.006         & 0.006         & 
\end{tabular}
\caption{\label{tab:tab_4}
Table reporting the confidence values for the estimate of the parameter
\(\eta_0\). The noise model is a
composition of a totally depolarising and an amplitude damping channel, which is a comprehensive Markovian and completely positive
noise model. The value of fidelity corresponds to the parameters \(0.05\) for the
depolarising and amplitude damping channel.
}
\end{table}

Each table should be read as follows.
If the noise is known or it is suspected
to have an average gate fidelity value
in one of the values on Tables~\ref{tab:tab_1}-\ref{tab:tab_4}
then, according to the confidence expected to obtain (we use 0.95, 0.999, and 1),
the number of shots and circuits required to obtain such confidence intervals
are displayed.
For concreteness, we explain the reading of Tab.~\ref{tab:tab_1}:
fixing the number of shots and circuits to 100 and for a noise with fidelity
\(0.89\),
the values resulting from the fitting of~Eq.~\eqref{eq:survival-probability-initialstate}
have an error of \(0.003\) with a frequency of \(0.95\),
error  \(0.006\) with frequency~\(0.999\), 
and  error \(0.008\) in any case.
This could also be written as 
\(\mathrm{Prob}(\abs{\eta_0-\check{\eta}_0}>0.003)  = 1-0.95\),
\(\mathrm{Prob}(\abs{\eta_0-\check{\eta}_0}>0.006)  = 1-0.999\),
and
\(\mathrm{Prob}(\abs{\eta_0-\check{\eta}_0}>0.008)  = 0\), respectively.

For the ququart case,
we get similar results, as presented in Tables~\ref{tab:tab_5}-\ref{tab:tab_7}:
\begin{table}[H]
  \centering
\begin{tabular}{llllll}
Shots & Circuits & \multicolumn{3}{l}{Error given a probability} & Fidelity              \\
 \(s\)     &  \(r\)         & 0.95          & 0.999         & 1             & \multirow{4}{*}{0.90} \\
      \hline
100   & 10       & 0.008         & 0.014         & 0.015         &                       \\
10    & 100      & 0.009         & 0.014         & 0.018         &                       \\
10    & 10       & 0.027         & 0.050         & 0.052         &                       \\
20    & 20       & 0.013         & 0.021         & 0.023         & 
\end{tabular}
\caption{\label{tab:tab_5}
Table reporting the confidence values for the estimate of the parameter. In contradistinction with the qutrit case, the noise
model corresponded to a randomly sampled channel with an average gate fidelity \(0.90\).
}
\end{table}

\begin{table}[H]
  \centering
\begin{tabular}{llllll}
Shots & Circuits & \multicolumn{3}{l}{Error given a probability} & Fidelity              \\
\(s\)      & \(r\)         & 0.95          & 0.999         & 1             & \multirow{4}{*}{0.99} \\
      \hline
100   & 10       & 0.001 & 0.002 & 0.002         &                       \\
10    & 100      & 0.001 & 0.002 & 0.002         &                       \\
10    & 10       & 0.004 & 0.006 & 0.008         &                       \\
20    & 20       & 0.002 & 0.003 & 0.004         & 
\end{tabular}
\caption{\label{tab:tab_6}
Table reporting the confidence values for the estimate of the parameter. In contradistinction with the qutrit case, the noise
model corresponded to a randomly sampled channel with an average gate fidelity \(0.99\).
}
\end{table}

\begin{table}[H]
  \centering
\begin{tabular}{llllll}
Shots & Circuits & \multicolumn{3}{l}{Error given a probability} & Fidelity              \\
 \(s\)   &    \(r\)      & 0.95          & 0.999         & 1             & \multirow{4}{*}{0.95} \\
      \hline
100   & 10       & 0.004 & 0.007 & 0.007        &                       \\
10    & 100      & 0.004 & 0.007 & 0.007        &                       \\
10    & 10       & 0.013 & 0.026 & 0.034        &                       \\
20    & 20       & 0.006 & 0.012 & 0.013        & 
\end{tabular}
\caption{\label{tab:tab_7}
Table reporting the confidence values for the estimate of the parameter. In contradistinction with the qutrit case, the noise
model corresponded to a randomly sampled channel with an average gate fidelity \(0.95\).
}
\end{table}

From the numerical evidence gathered above we see that the number of shots and
random circuits required is around 20. 
Our numerical results suggest that doubling the shots and circuits
exponentially decreases the error.

\begin{table}[H]
  \centering
\begin{tabular}{llll}
\hline
Strategy      & \multicolumn{3}{l}{Error given a probability} \\
circuit depth & 0.95         & 0.999        & 1           \\ \hline
i             & 0.002       & 0.003       & 0.004       \\ 
ii            & 0.002       & 0.003       & 0.004       \\ 
iii           & 0.003       & 0.004       & 0.006       \\ 
iv            & 0.001       & 0.001       & 0.002       \\ 
\end{tabular}
\caption{\label{tab:table-strategies}
  Comparison of the strategies.
  The result is based on the confidence intervals.
  For the noise we used a depolarising channel composed with a phase-damping;
  the depolarising and dephasing parameters are set to \(0.1\).
}
\end{table}

In this paragraph,
we discuss the best circuit depth strategy to obtain smaller errors in the estimate parameters \(\eta_{\IrrepIndex}\).
The strategy is drawn from the results presented in~Table~\ref{tab:table-strategies}.
First,
fix 
the noise to be a composition of a depolarising and phase-damping channels,
which amounts for a comprehensive kind of noise present in current
implementations.
Then,
estimate the parameter \(\eta_0\) according to certain chosen circuit depths.
The strategies chosen are:
\begin{enumerate*}
  \item \(m \in \{2,4,8,16,32,64\}\).
  \item \(m \in [8]\),
  \item \(m \in [5]\),
  \item spaced values \(m \in \{5,10,15,20,\ldots, 100\}\).
\end{enumerate*}
The best strategy, 
resulting from the simulation and reported on Table~\ref{tab:table-strategies},
is taking multiples of~\(5\) (strategy iv).
The order of strategies is 
iv,
i,
ii,
iii.
Changing the spacing in strategy~iv did not improve the result significantly.

We finish this section with an explicit construction of the group elements for
the qutrit and ququart versions.
First we discuss the qutrit and then the ququart.
The qutrit case is simpler since \(S_3\) representatives can be simply stated.
Let \(\RepSymm(\sigma)\) be the representative of some permutation.
Then the two generators of the diagonal unitary matrices that appear in
\thegroup\ are:
 \begin{equation}
T_0 \coloneqq  \Diag[
\RootUnity_9^{2},
\RootUnity_9^{4},
\RootUnity_9^{3}
]
\text{ and }
T_1 \coloneqq  \Diag[
\RootUnity_9^{2},
\RootUnity_9^{2},
\RootUnity_9^{5}
].
\end{equation}
Thus an arbitrary group element is
\begin{equation}
\RepSymm(\sigma)T_0^{\alpha_0}
T_1^{\alpha_1},
\end{equation}
where \(\alpha_0, \alpha_1 \in [9]\). 
Both \(T_0\) and \(T_1\) have order \(9\);
the \thegroup\ has order  \(9^2\times 6 = 486\).

The ququart case is more interesting due to the changing Heisenberg-Weyl
matrices for the following tensor product of qubit matrices
\begin{equation}
  \langle
\sigma_x\otimes \identity,
\sigma_z\otimes \identity,
\identity\otimes \sigma_x,
\identity\otimes \sigma_z
 \rangle,
\end{equation}
where \(\sigma_x\) and \(\sigma_z\) and the  \(X\) and \(Z\) matrices for
qubits.
The T gate is \(T\times \identity\) with~\(T = \Diag[1, \omega_{8}]\).

\section{Discussion}
In this section, we provide an overview of our results and analysis.
We introduced a scheme for characterising a qudit gate set (in powers of prime
dimension) that includes a non-Clifford generator of a universal gate set.
Our scheme is feasible (as it both requires three single-qudit and
one control primitive gates) and scalable (as the number of parameters to
estimate remains two, independent on the dimension of the system).
Also our method is practical
as it requires equal or fewer resources compared
to the resources needed for Clifford randomised benchmarking.
Additionally, our scheme can be used to characterise a qudit T gate.
Only two parameters are needed to estimate the average gate fidelity.

Our method applied to qubits reduces to the qubit dihedral benchmarking scheme,
which has the nice feature of a bi-parametric average gate fidelity.
These parameters can be estimated using the techniques of a standard RB experiment.
The required gate set for our method has fewer gates than for the Clifford gate set.

These features come as a result of leveraging methods from representation
theory,
ring theory, quantum channels, and combinatorics.
We used representation theory and ring theory to construct a group we call rHDG.
This group has a unitary irrep (unirrep) 
that
transforms---by twirling---any channel 
into a bi-parametric channel.
From this transformation, we can recover expressions for gate fidelity and sequence fidelity.

Our proof that the twirl of a channel by the rHDG unirrep is bi-parametric,
hinges upon a 
combinatorial identity given in Eq.~\eqref{eq:novel-identity}.
This identity relates the number of partitions of a finite set (Bell numbers) with a sum over the partitions of an integer.

We identified some unexpected outcomes.
The Abelian subgroup of the rHDG varies structurally for 
systems with different numbers of states.
To efficiently construct a unirrep of the rHDG, we use Howell's algorithm~\cite{Howell1986}.
We observed an unexpected behaviour for Clifford gates:
the qubit Clifford gate corresponding to the transition~\(\ket{0} \leftrightarrow \ket{1}\)
is not a Clifford operation for~\(\qudim>3\) level systems;
as a matter of fact, it is not a member of any level of the qudit Clifford
hierarchy.

We proceed to further explain our feasibility claim.
Our scheme requires the estimate of two parameters,
which translates to the requirement of two initial states.
Each sequence fidelity is a single exponential-decaying function.
The number of parameters in our scheme is equal to the number of different initial states required.
Consequently, our scheme requires twice---compared to Clifford randomised
benchmarking---the number of experiments 
for a simple data analysis.
Additionally, we have proven that, for any number of qudits,
the states~\(\ket{0}\) and~\(H\ket{0}\) provide access to the two distinct parameters necessary for fidelity estimation.

Our scheme is the first to address non-Clifford qudit gates.
Therefore the only valid comparison is against
qudit Clifford randomised benchmarking.
Our scheme demands similar or fewer experimental resources,
although the number of experiments might need to be doubled for thorough validation.
We are now prepared to conclude.

\section{Conclusion}

We presented our scheme to completely characterise a universal multi-qudit gate
set. 
The characterisation is done by introducing a gate set that includes a qubit T
gate.
Our scheme is feasible, as it requires the estimation of only two parameters,
irrespective of the level system (qudit) and the number of qudits.
Additionally,
as our scheme requires the magic T gate, X, and~\(X_{01}\) gates, 
and as the multiplication rule of such elements is easier to compute than for the Clifford case,
our scheme is simpler to implement than Clifford RB.
Moreover, by our qutrit extension of non-Clifford
interleaved benchmarking, we demonstrated that our gate-set (the rHDG) is a test~bed for extending RB
schemes beyond the unitary 2-design assumption.
Our method paves the way for a complete characterisation of a universal
gate set, and could be used in current platforms that made use of IB to
characterise non-Clifford gates~\cite{Rengaswamy_Calderbank_Pfister_2019,
Morvan_Ramasesh_Blok_Kreikebaum_OBrien_Chen_Mitchell_Naik_Santiago_Siddiqi_2021,Hrmo2023,Goss2022}.

Currently, there are multiple qudit implementations, spanning from qutrits to quoctits
(systems with up to eight levels)~\cite{Wang2020,
Chi2022,Imany2019,Lanyon2009,min2018,Malik2016,Lanyon_Weinhold_Langford_OBrien_Resch_Gilchrist_White_2008, 
randall2015,leupold18,Klimov_Guzman_Retamal_Saavedra_2003,Zhang2013, Hrmo2023,
Luo2023,Seifert2023,Roy2022,Kononenko_Yurtalan_Ren_Shi_Ashhab_Lupascu_2021, Morvan_Ramasesh_Blok_Kreikebaum_OBrien_Chen_Mitchell_Naik_Santiago_Siddiqi_2021,Goss2022, 
fernandez2022coherent,
Lindon2022, 
Fu2022,guo2023}.
Several of these implementations employ randomised benchmarking to characterise
their Clifford
operations~\cite{Ringbauer2022,Morvan_Ramasesh_Blok_Kreikebaum_OBrien_Chen_Mitchell_Naik_Santiago_Siddiqi_2021,Seifert2023}.
Our work broadens the scope of this characterisation, enabling these systems to assess a universal gate set comprehensively.
Furthermore, a subset of these implementations incorporates controlled operations.
Consequently, our scheme's scalability is particularly interesting to
characterise multi-qudit universal gate sets,
allowing the scalable of any implementation of a universal gate set.

  As our scheme deals with multi-qudit gates, 
it is natural to ask about the detection of cross-talk errors.
Because extending the techniques employed in the detection of cross-talk errors
requires knowledge of the irrep decomposition of~\(\thegroup\otimes\thegroup\)~\cite{GambettaCorcoles2012},
we leave this extension for future work.
However, extension of detection of cross-talk errors first should be done using the
Clifford group;
we anticipate this analysis to be more challenging than the qubit
case~\cite{Wallman_Flammia_2014,GambettaCorcoles2012}.

\ack
The work of HdG and BCS is supported by the Natural Sciences and Engineering Research Council of Canada and by the Government of Alberta.
\bibliography{manuscript}
\appendix
\numberwithin{equation}{section}
\renewcommand{\thesection}{\Alph{section}}
\section{Construction of group elements}
In this appendix, we deal with the construction of the representation for group
elements.
We present three constructions, in increasing order of sophistication.
The first two are presented here and the last one is discussed entirely in the
next appendix due to its complexity.

The first construction is for the largest group that contain a T gate.
It includes the permutation matrices and every permutation of the diagonal entries
of a matrix of the form
\begin{equation}
D_1(\Order(\qudim)) \coloneqq 
\Diag[\omega_{\Order(\qudim)}, 1,\ldots, 1],
\end{equation}
where \(\RootUnity_\qudim = \exp(2\pi\myi/\qudim)\).
In general \(D_i\) has \(\RootUnity_\qudim\) in the \(i\)-th position of the
diagonal and the rest are \(1\)s.

Thus, given a permutation \(\sigma\) and a subset \(J\) of \([\qudim]\) a group
element of \thegroup\ is of the form
\begin{equation}
\RepSymm(\sigma) \prod_{i\in J} D_i.
\end{equation}
This concludes the construction of the maximal \thegroup;
two examples (for qutrit and ququart systems) are constructed in
\S\ref{sec:numerics}.

The next construction can be considered as a redundant construction.
The following construction is not optimal since 
it introduces
terms that are product of other gates.
However, the construction is simple. We deal with a non-redundant construction
in Appendix~\ref{app:product-cyclic-groups}.

Let \(T\) be the diagonal gate that is to be characterised.
We can write it as \(T = \sum_i \delta_{i,i} \omega^{J_i}\),
where \(J_i\in [\Order(\qudim)]\) are the powers of \(\RootUnity_{\Order(\qudim)}\).

\section{Identifying the product of cyclic groups}\label{app:product-cyclic-groups}
\thegroup\ is a semidirect product of a symmetric group
and a product of cyclic groups.
This product of cyclic groups depends on the matrix \(T\).
In this appendix, we clarify 
the process to identify the product of cyclic groups.
This appendix is relevant for \S\ref{sec:approach}.

We start our discussion---about the identification of the product of cyclic
groups---by introducing notation.
Let \(V \coloneqq \toRing(T)\), where 
\(\toRing\) denotes the mapping that extracts the diagonal of a square matrix,
then computes the argument of the entries and divides them by
\(2\pi/\Order(\qudim)\).
As an example of the usage of \(\toRing\),
recall 
\(T = \Diag[1,\RootUnity_9,\RootUnity_9^{8}]\)
for qutrits; we get \(\toRing(\Diag[1,\RootUnity_9,\RootUnity_9^{8}]) =
[0,1,8]\).

We now define the data structure used in the algorithm to identify the
generators of the cyclic groups.
Remembering \(T_\Permutation = 
\RepSymm({\Permutation})
T
\RepSymm({\Permutation})^{\dagger}\) and \(\bm{T}' \coloneqq \langle 
T_\Permutation
\colon \Permutation\in S_{\QuditDim} \rangle\)---given in Eq.~\eqref{eq:def-permutations-t-gate}---from \S\ref{sec:approach}, 
we define the multiset \(\bm{V} \coloneqq  \{\toRing(T_\Permutation)\colon T_\Permutation
\in \bm{T}'\}\).
Each \(\toRing(T_\Permutation)\) in \(\bm{V}\) is arranged as a column into a
\(d\times d!\) matrix \(M\).
This matrix has entries
\begin{equation}\label{eq:entries-matrix-m}
 M_{\Permutation,j} \coloneqq \toRing(T_\Permutation)_j,
\end{equation}
where we use lexicographic ordering to list the set of permutations.

We provide an example on how to list a sequence of permutations for a set with
three elements.
Consider two vectors, denoted as \(v\) and \(v'\), each having entries from the set~\([3]\),
with individual entries represented by \(v_i\) and \(v_i'\), respectively.
The criterion for establishing \(v > v'\) is dictated by the condition that,
for all indices \(i\) ranging from \(0\) to a specified non-zero index, \(v_i \geq v_i'\) holds true.
The implementation of this ordering on permutations is achieved by examining the
permutation's effect on the ordered set~\([3]\).
By using this method, we can list the permutations in \(S_3\), resulting in the ordered sequence \(\{e, (23), (12), (132), (123), (13)\}\).

The set \(\Span(\bm{V})\) is isomorphic to \(\bm{T}'\)---addition for vectors corresponds to matrix multiplication,
and scalar multiplication in \(\Span(\bm{V})\) is computing a power of the
matrix in \(\langle \bm{T}'\rangle\).
In particular, the notion of the  order of matrix is then translated into the order of a
vector.
The order of a vector \(v\) with entries in some ring \(R\) is
the non-zero ring element \(r\neq 0\in R\) that
\(rv = \bm{0}\in \Span(\bm{V})\).

We now explain how to
identify
the product of cyclic groups that is isomorphic to 
\(\bm{T}'\).
We compute a basis for \(\Span(\bm{V})\);
the order of the elements of the basis  is equal to the order of one of the product of the
cyclic group isomorphic to 
\(\bm{T}'\).
For a single- and multi-qubit system, this isomorphism reduces to the one used
in multi-qubit dihedral benchmarking~\cite{Cross2016}.
The computation of the basis is done by 
computing the row-echelon form of \(M\) in Eq.~\eqref{eq:entries-matrix-m}~\cite{Howell1986}.

We now consider an example relevant for the qutrit case.
The process is different than that encountered in linear algebra textbooks as the
entries of \(M\) are members of a ring.
Therefore, we use Howell's algorithm to compute the row-echelon form~\cite{Howell1986}.
We start by identifying \(T\), then computing \(M\).
Starting with \(T = \Diag[1,\RootUnity_9,\RootUnity_9^{8}]\),
the matrix \(M\) is
\begin{equation}
M = 
\begin{bmatrix}
 0 & 0 & 1 & 1 & 8 & 8 \\
 1 & 8 & 0 & 8 & 0 & 1 \\
 8 & 1 & 8 & 0 & 1 & 0 
\end{bmatrix},
\end{equation}
where each entry is a member of \(\Ring_9\).
The row-echelon form of \(M\) is
\begin{equation}\label{eq:row-echelon}
 \operatorname{row-echelon}(M)
 =
 \begin{bmatrix}
 1 & 8 & 0 & 8 & 0 & 1 \\
 0 & 0 & 1 & 1 & 8 & 8 \\
 0 & 0 & 0 & 0 & 0 & 0 
 \end{bmatrix}.
\end{equation}
As in the complex case,
the columns of the leading units in \(\operatorname{row-echelon}(M)\)
correspond to the indices, in \(\bm{V}\), of the basis vectors.
Here, these leading units are in columns 1 and 3.
Then---by the output of Howell's algorithm in Eq.~\eqref{eq:row-echelon}---the
basis for \(\Span(M)\) is the ordered set \(\{(0,1,8), (1,0,8)\}\),
obtained by keeping columns~1 and 3 of the matrix \(M\).
Therefore, by using the isomorphism between matrices and vectors,
\(\bm{T}' \cong \langle \Diag[1,\RootUnity_9,\RootUnity_9^{8}],
\Diag[\RootUnity_9,1,\RootUnity_9^{8}]\rangle \cong C_9\times C_9\).
\section{Properties of the qudit rHDG relevant for Randomised Benchmarking schemes}
\label{appendix:properties-rhdg}
\subsection{Background}

In this subsection we discuss background material needed for the technical proofs of our scheme.
We review properties of the theory of characters for finite groups and the Pauli-Liouville representation.

Let $\FiniteGroup$ be a finite group and $\GroupElement$ an element of $\FiniteGroup$.
Let $\Irrep$ be a representation of $G$.
Then the character of the element $\GroupElement$ in irrep $\Irrep$ is denoted by $\Character_{\Irrep}(\GroupElement)$.
If $\GroupElement$ is labelled by some set of labels $\bm{L}$, as in for example a direct product, we denote the character
of $\GroupElement$ by $\Character_{\Irrep}(\bm{L})$.

We recall the following fact about the inner product of the character of some
representation with itself.
\begin{lemma}[Adapted from~\cite{serre1977}]\label{thm:number-irrpes}
 Let $\FiniteGroup$ be a finite group.
 Let $I$ be a finite index-set and $\{\Irrep_i\}_{i\in I}$ be a finite list of irreps of
 $\FiniteGroup$.
Suppose a representation $\Irrep' = \bigoplus_{i\in I} k_i \Irrep_i$; that is,
$\Irrep'$ is a representation with $|I|$ distinct irreps, each irrep $\Irrep_i$ appearing
$k_i$ times.
Let $\Character_\Irrep(\GroupElement)$ be the character of the element $\GroupElement\in \FiniteGroup$ in the representation $\Irrep'$.
Then
\begin{equation}
|\FiniteGroup|^{-1}\sum_\GroupElement
|\Character_{\Irrep'}(\GroupElement)|^2
= \sum_i k_i^2.
\end{equation}
In other words,
the quantity \(\frac{1}{|\FiniteGroup|}\sum_\GroupElement|\Character_{\Irrep'}(\GroupElement)|^2\)
returns the sum of the squares of the frequency of each irrep included in the
representation~\(\Irrep'\).
\end{lemma}

The second important concept is the Pauli-Liouville representation of a quantum channel \(\Channel_{\SetOfKraus}\).
Consider a set of Kraus operators \(\SetOfKraus\) defined in
Eq.~\eqref{eq:definition-condition-kraus-operator}. The Pauli-Liouville (PL)
representation of the channel \(\channel_{\SetOfKraus}\)---defined in
Eq.~\eqref{eq:kraus}---is the matrix
\(\RepPlRep(\SetOfKraus)\) with entries
\begin{equation}
\RepPlRep(\SetOfKraus)_{i,j}
\coloneqq 
\frac{1}{\sqrt{\QuditDim}}\trace(W_i^{\dagger} \sum_k A_k W_jA_k^{\dagger});
\end{equation}
\(W_i\) is a HW-matrix defined in Eq.~\eqref{eq:hw-matrix}.
As the set \( \SetOfKraus\) defines a channel \(\Channel\), we also use
\(\RepPlRep(\Channel)\) to denote the PL representation of a channel \(\Channel\).
Similarly, for a unitary matrix~\(U\), we denote by \(\RepPlRep(U)\) the PL
representation for \(\SetOfKraus = \{U\}\).

\subsection{Number of irreps in the Pauli-Liouville representation}

In this subsection we show that the PL representation of the \thegroup---defined in
Eq.~\eqref{eq:the-group}---decomposes into three irreps.
We introduce notation used in other parts of this section.
We conclude with the proof of the tri-partite decomposition of the PL representation,
which is the mathematical result at the root of our decomposition given in
Eq.~\eqref{eq:irrep-decomposition}.

For later convenience, we start by introducing an auxiliary group.
Consider the set of matrices
\begin{equation}
\bm{D}_{\QuditDim} \coloneqq 
\{
 \Diag[\RootUnity_{\Order(\QuditDim)},1,\dots,1],
 \ldots,
 \Diag[1,\dots,1,\RootUnity_{\Order(\QuditDim)}]
\},
\end{equation}
where \(\Order(\qudim)\) is defined in Eq.~\eqref{eq:order-t-gate}.
Notice
\begin{equation}
\langle \bm{D}_\QuditDim\rangle \cong C_{\Order(\QuditDim)}^{ \QuditDim},
\end{equation}
where we use the brackets to mean ``generated''~\cite{MikeAndIke}.

Additional notation is required.
Consider a permutation \(\Permutation\in S_{\QuditDim} \) acting on
\([\qudim]\); let~\(f_\Permutation\) denote the
cardinality of the set of stable points under the action of~\(\Permutation\).
Similarly, consider the conjugacy class---in \(S_\qudim\)---labelled by 
the integer partition $\Partition\vdash \QuditDim$.
Every element in the conjugacy class \(f_\Partition\) leaves invariant
the same number of points;
we denote by \(f_\Partition\) the number of fixed points for any
\(\Permutation\in\Partition\).
We denote
the number of permutations in \(\Partition\) by
\(c(\Partition)\).

\begin{lemma}\label{lemma:trivial-identity}
For any \(k\in \Ring_{\Order(\QuditDim)}^{\QuditDim}\), we have 
\begin{equation}
\sum_{\CyclicElement\in \Ring^{\QuditDim}_{\Order(\QuditDim)}}
\RootUnity_{\Order(\QuditDim)}^{k \CyclicElement_i} = 0.
\end{equation}
\end{lemma}
\begin{proof}
  The sketch of the proof goes as follows.
This result is equivalent to the sum over the roots of the identity.
This sum is always zero.
\end{proof}
\begin{lemma}[Partial character of the rHDG]\label{lemma:sum-over-alpha}
 Let \(\Permutation\in S_{\QuditDim} \) and \(f_\Permutation\) denotes the cardinality of the set of points fixed by \(\Permutation\).
 Furthermore, recall
 that elements of \thegroup\ are labelled by \((\Permutation,\CyclicElement)\)
 as per Eq.~\eqref{eq:representation-gamma-unirrep-rhdg}.
 Then, for $\Permutation$,
\begin{equation}
|\bm{D}_{\QuditDim}|^{-1}
\sum_{\CyclicElement\in C_{\Order(\QuditDim)}^{\QuditDim}}
\Character_\RepPlRep(\Permutation,\CyclicElement)^2
=
\Average_{\CyclicElement\in C_{\Order(\QuditDim)}^{\QuditDim}}
\Character_\RepPlRep(\Permutation,\CyclicElement)^2
= f_\Permutation (2f_\Permutation - 1),
\end{equation}
where \(\Character_\RepPlRep(\Permutation,\CyclicElement)\)
is the character of the PL representation of \thegroup.
\end{lemma}
\begin{proof}
Recall \(\UniRepHdg\) is the unirrep of \thegroup defined in
Eq.~\eqref{eq:representation-gamma-unirrep-rhdg}.
Let \(\Character_\UniRepHdg(\Permutation,\CyclicElement)\) be the character of the element
\(\RepSymm(\Permutation) \RepCyclic(\CyclicElement)\in \Range\UniRepHdg\), 
where \(\RepCyclic(\CyclicElement) = \Diag[\RootUnity_{\Order(\QuditDim)}^{\CyclicElement_1},\ldots,\RootUnity_{\Order(\QuditDim)}^{\CyclicElement_{\QuditDim} }]\)
and \(\RepCyclic\) is defined in Eq.~\eqref{eq:def-representation-symmetricgroup}.
Then 
 \begin{equation}
\Character_\UniRepHdg(\Permutation,\CyclicElement)
\coloneqq
\trace(
\RepSymm(\Permutation) \RepCyclic(\CyclicElement)
) = \sum_{i\in J} \RootUnity^{\CyclicElement_i}_{\Order(\QuditDim)},
\end{equation}
where \(J\) is the index-set for the positions of the non-zero entries of the
diagonal of~\(\RepSymm(\Permutation)\)
and the cardinality of \(J\) satisfies \(|J| = f_\Permutation\).

From \(\chi_{\gamma}\) the character of the Pauli-Liouville representation:
\begin{subequations}
\begin{equation}\label{eq:standard-irrep-in-plrep}
  \chi_\Gamma(\sigma,\alpha)
  =
  \sum_{i,j\in [J(\sigma)]}\omega_{\orderof(d)}^{\alpha_i-\alpha_j}
  =
  f_\sigma+\sum_{i\neq j}\omega_{\orderof(d)}^{\alpha_i-\alpha_j}
\end{equation}
note that \(\chi_\Gamma(\sigma,\bm{\alpha})\) is a real number,
thus we omit the modulus appearing in Theorem~\ref{thm:number-irrpes}.
The next step is averaging \({\chi_\Gamma}^2\) (the square of
\(\chi_\Gamma(\sigma,\bm{\alpha})\)) over every~\thegroup\ element and
verifying that
\(\average_{\sigma,\bm{\alpha}}{\chi_\Gamma(\sigma,\bm{\alpha})}^2 = 3\).
Then, using Theorem~\ref{thm:number-irrpes}, show that there are only three inequivalent
irreps in the \thegroup.

First, we compute \(\chi^2_\Gamma(\sigma,\alpha)\):
\begin{align}
\chi^2_\Gamma(\sigma,\alpha)
&=
(f_\sigma + \sum_{i\neq j}\omega_{\orderof(d)}^{\alpha_i-\alpha_j})
(f_\sigma + \sum_{u\neq v}\omega_{\orderof(d)}^{\alpha_u-\alpha_v})
\\&=
f_\sigma^2 + 
f_\sigma (\sum_{i\neq j}\omega_{\orderof(d)}^{\alpha_i-\alpha_j})+
f_\sigma (\sum_{u\neq v}\omega_{\orderof(d)}^{\alpha_u-\alpha_v})
+ (\sum_{i\neq j,u\neq v}\omega_{\orderof(d)}^{\alpha_i-\alpha_j+\alpha_u-\alpha_v})
\end{align}

We note that, for \(i\neq j\)
\begin{equation}
\average_{\bm{\alpha}}
\omega_{\orderof(d)}^{\alpha_i-\alpha_j} = 0.
\end{equation}
Thus
\begin{equation}\label{eq:rest-average}
  \average_{\bm{\alpha}} \sum \chi^2_\Gamma(\sigma,\bm{\alpha})
=
f_\sigma^2+
\average_{\bm{\alpha}} \sum_{i\neq j,u\neq v}\omega_{\orderof(d)}^{\alpha_i-\alpha_j+\alpha_u-\alpha_v}
\end{equation}
To simplify Eq.~\eqref{eq:rest-average} we need to isolate the phases
\(\alpha_i\) appearing in the exponent of \(\rootunit\).
Only then we can compute the average over the phases.
We decompose the cases \(i\neq j\) and \(u\neq v\) in
Table~\ref{tab:phase-classification}.

\begin{table}[h]
  \label{tab:phase-classification}
  \caption[Table with phases in the proof]{Table with the clasification of the
  phases of Eq.~\eqref{eq:rest-average}.}
  \centering
\begin{tabular}{l|l|l|l|l}
\multicolumn{4}{c}{Configuration} & $\alpha_i-\alpha_j+\alpha_u-\alpha_v$ \\
\hline
 \(i\neq v\)  &  \(i=u\)  & \(j\neq u\)   & \(j=v\)   & \(2\alpha_i-2\alpha_j\)  \\
 \(i\neq v\)  &  \(i=u\)  & \(j\neq u\)   & \(j\neq v\)   & \(2\alpha_i-\alpha_j-\alpha_v\)  \\
 \(i=v\)  &  \(i\neq u\)  & \(j=u\)   & \(j\neq v\)   & \(0\)  \\
 \(i\neq v\)  &  \(i\neq u\)  & \(j= u\)   & \(j\neq v\)   & \(\alpha_i-\alpha_v\)  \\
 \(i\neq v\)  &  \(i\neq u\)  & \(j\neq u\)   & \(j = v\)   & \(\alpha_i+\alpha_u-2\alpha_v\)  \\
 \(i=v\)  &  \(i\neq u\)  & \(j\neq u\)   & \(j\neq v\)   & \(\alpha_u-\alpha_j\)  \\
 \(i\neq v\)  &  \(i\neq u\)  & \(j\neq u\)   & \(j\neq v\)   & \(\alpha_i-\alpha_j+\alpha_u-\alpha_v\)  \\
\end{tabular}
\end{table}

On Table~\ref{tab:phase-classification} 
we describe the different combinations of phases that appear in the sum
Eq.~\eqref{eq:rest-average}.
We decompose the sum-average in the
second summand in the right-hand side of~Eq.~\eqref{eq:rest-average}:
\begin{align}
\average_{\bm{\alpha}} \sum_{i\neq j,u\neq v}\omega_{\orderof(d)}^{\alpha_i-\alpha_j+\alpha_u-\alpha_v}
&=
\average_{\bm{\alpha}}
\sum_{\substack{i\neq v, i= u,\\j\neq u, j\neq v}} \rootunit_{\orderof(\qudim)}^{2\alpha_i-2\alpha_j}
+
\average_{\bm{\alpha}}
\sum_{\substack{i\neq v ,i=u,\\ j\neq u, j\neq v}} \rootunit_{\orderof(\qudim)}^{2\alpha_i-\alpha_j-\alpha_v}
\\&\quad+
\average_{\bm{\alpha}}
\sum_{\substack{i=v,i\neq u,\\ j=u, j\neq v}} \rootunit_{\orderof(\qudim)}^{0}\label{eq:amlo}
\\&\quad+
\average_{\bm{\alpha}}
\sum_{\substack{i\neq v,i\neq u,\\j= u,j\neq v}} \rootunit_{\orderof(\qudim)}^{\alpha_i-\alpha_v}
+
\average_{\bm{\alpha}}
\sum_{\substack{i\neq v,i\neq u,\\j\neq u,j = v}} \rootunit_{\orderof(\qudim)}^{\alpha_i+\alpha_u-2\alpha_v}
\\&\quad+
\average_{\bm{\alpha}}
\sum_{\substack{i=v,i\neq u,\\j\neq u,j\neq v}} \rootunit_{\orderof(\qudim)}^{\alpha_u-\alpha_j}
+
\average_{\bm{\alpha}}
\sum_{\substack{i\neq v,i\neq u,\\j\neq u,j\neq v}} \rootunit_{\orderof(\qudim)}^{\alpha_i-\alpha_j+\alpha_u-\alpha_v}.
\end{align}
Note that each sum, except \eqref{eq:amlo},
has in the exponent a sum of phases, with each phase different.
Thus, averaging over the phases we obtain zero.
Only the sum~\eqref{eq:amlo} is non-zero:
\begin{equation}
\average_{\bm{\alpha}}
\sum_{\substack{i=v,i\neq u,\\ j=u, j\neq v}} \rootunit_{\orderof(\qudim)}^{0} 
=
\sum_{\substack{i=v,i\neq u,\\ j=u, j\neq v}} 1
= f_\sigma(f_\sigma-1).
\end{equation}
Thus,
\begin{equation}\label{eq:last-step-phases}
\average_{\bm{\alpha}} \sum \chi^2_\Gamma(\sigma,\alpha) = f^2_\sigma + f_\sigma(f_\sigma-1)
=
2f_\sigma^2-f_\sigma.
\end{equation}
\end{subequations}
Note that the computations needed to arrived at the last equation (Eq.~\eqref{eq:last-step-phases})
are valid for any phase order greater than two.
This shows our scheme is valid for any diagonal gate with order greater than two.
\end{proof}

We now recall 
an identity for the moments over a distribution on partitions~\cite{rota1964}.
Consider a finite set \(\finiteset\) with \(\numberlements\) elements
\(\finiteset \coloneqq \{\setelement_0,
\ldots, \setelement_{\numberlements-1}\}\).
The number of disjoint partitions of \(\finiteset\) is called the
\(\numberlements\)-th Bell number; this number is denoted by \(B_\numberlements\).
\begin{theorem}[Relation between integer partitions and Bell numbers~\cite{rota1964}]\label{thm:bellidentity}
 Let \(\numberlements\) be a positive integer and \(B_\numberlements\) the
 \(\numberlements\)-th Bell
 number.
 Let \(c(\lambda)\) be the number of permutations with cycle decomposition \(\lambda\).
 For a given integer partition \(\Partition\vdash \numberlements\), let
 \(c(\Partition, \auxdummyindex)\)
 denote the number of times \(\auxdummyindex\) appears in \(\Partition\).
 Then
 \begin{equation}\label{eq:novel-identity}
B_\numberlements =
\sum_{\Partition\vdash \numberlements}
\frac{c(\Partition,1)^{\numberlements} }{\prod_\auxdummyindex^{\numberlements}
\auxdummyindex^{c(\Partition,\auxdummyindex)} c(\Partition, \auxdummyindex)}.
 \end{equation}
Note that 
\(c(\Partition,1)\)
is the number of fixed points in a permutation corresponding to the class
labelled by \(\Partition\); that is, \(c(\Partition,1) = f_\sigma\) if  \(\sigma\) is a
permutation with cycle decomposition \(\Partition\).
\end{theorem}
\begin{theorem}[rHDG PL-representation tripartite decomposition]\label{thm:decomposition_plrep}
The PL representation decomposes into three irreps with trivial 
multiplicity (multiplicity equal to one).
\end{theorem}
\begin{proof}

 \begin{subequations}
\begin{align}
  \frac{1}{|\thegroup|}
  \sum_{\Permutation,\CyclicElement\in\thegroup}
| \Character_\RepPlRep(\Permutation,\CyclicElement)|^2
 &=
 \frac{1}{\QuditDim!}
 \frac{1}{|D_{\QuditDim} |}
 \sum_{\Permutation\in S_{\QuditDim} ,
 \CyclicElement\in C_{\Order(\QuditDim)}^{\QuditDim}}
 \Character_\RepPlRep(\Permutation,\CyclicElement)^2
 , 
 \\
 &=
 \frac{1}{\QuditDim!}
 \sum_{\Permutation\in S_{\QuditDim} }
 f_\Permutation(2 f_\Permutation-1), \\
 &=
 \frac{1}{\QuditDim!}
 \sum_{\Partition\vdash \QuditDim}
 f_\Partition(2 f_\Partition-1) c(\Partition),\\
 &=
 \sum_{\Partition\vdash \QuditDim}
 \frac{c(\Partition,1)(2c(\Partition,1)-1)}{\Projector_{\auxdummyindex=1}^{\QuditDim}
 \auxdummyindex^{c(\Partition,\auxdummyindex)} c(\Partition,\auxdummyindex)!} \\
 &=2B_2-B_1= 3.
\end{align}
 \end{subequations}

Let $n_i\in\integers$ for $i\in[3]$.
Then the only solution to the equation
$\sum n_i^2 = 3$ is 
to have three non-zero \(n_i\) and those non-zero  are equal to  \(1\).
Therefore, by Theorem~\ref{thm:number-irrpes}, there are three inequivalent
irreps in the decomposition of the Pauli-Liouville representation
of the rHDG.
\end{proof}

\subsection{Identification of the irreps}
From the proof of Thm.~\ref{thm:decomposition_plrep},
we know that there are three inequivalent irreps in the PL representation of the
\groupname{}.
We now proceed to identify them.

In 
Eq.~\eqref{eq:standard-irrep-in-plrep}
notice that 
\begin{equation}
 \Character_{\RepPlRep}(\Permutation,\CyclicElement)
 =
|\Character_{\UniRepHdg}(\Permutation,\CyclicElement)|^2 = f_\Permutation + \Character_{D_{\QuditDim} }(\CyclicElement),
\end{equation}
where $\Character_{D_{\QuditDim} }(\CyclicElement)$ is a quantity to be determined after computing the trace of
a \groupname{} element.
Now, $f_\Permutation$ is the character of the standard representation of $S_{\QuditDim} $, the
representation obtained by permuting the rows of the \(\qudim\times\qudim\) identity matrix.

The standard representation of \(S_\qudim\) is defined by its action on a basis.
The action of this representation is by permuting the entries of a vector;
the permutation applied is \(\Permutation\in S_\qudim\).
Because there is a non-trivial invariant subspace of the standard
representation, 
namely the vector space corresponding to those vectors with each entry
equal, it is reducible.
However, the representation obtained by restricting to the orthogonal complement
of \(\Span\{(1,\dots,1)\}\), denoted by \(\Span\{(1,\dots,1)\}^{\bot}\),
is irreducible---this irrep is called the standard irrep of \(S_\qudim\).

Thus, we can decompose the carrier space of the 
reducible standard representation of \(S_\qudim\) into two irreps: the trivial
irrep and the standard irrep.
Next, notice that for \(\Permutation\in S_\qudim\), its character in the
standard representation is equal to  \(f_\Permutation\).
Thus, for any permutation \(\Permutation\), we can partition the 
integer \(f_\Permutation\) as
 \begin{equation}\label{eq:character-standard-irrep-s-qudim}
f_\Permutation
=
 \Character_\mathbb{I}+\Character_\labelirrepstd(\Permutation);
\end{equation}
the trivial character \(\Character_\identity\) 
corresponds to the 
one-dimensional invariant subspace spanned by \((1,\dots, 1)\).
Next we use Eq.~\eqref{eq:character-standard-irrep-s-qudim} 
to identify the irreps in the Pauli-Liouville representation of the \thegroup.

From the character in Eq.~\eqref{eq:character-standard-irrep-s-qudim}
and a simple computation using
the orthogonality theorem of characters~\cite{serre1977},
we obtain the following partition of the character of an element of the
\groupname{}
in the PL representation:
\begin{equation}
 \Character_{\RepPlRep}(\Permutation,\CyclicElement) =
 \Character_\mathbb{I}+\Character_\labelirrepstd(\Permutation) + \Character_{D_{\QuditDim} }(\CyclicElement).
\end{equation}
We use this decomposition to identify the irreps in the
Pauli-Liouville~representation.

\subsection{States used to compute the parameters of the sequence fidelity via
fitting}

In general, using an arbitrary initial state leads to sequence fidelity
expressible as a sum of
two exponential functions~\cite{helsen22prx}; this sum leads to a more complicated
data-analysis than for a single exponential function.
In this subsection we show 
how to simplify the data-analysis by reducing to a single exponential
decay; the resulting sequence fidelity is written in Eq.~\eqref{eq:survival-probability-initialstate}.
This simplification is achieved by using two different initial states in the
randomised benchmarking protocol.
We compute the projectors onto each different irrep in the tri-partite decomposition of the PL-representation.
Then we show how the state \(\ket{0}\) is mapped to the null vector by the
projector \(\Projector_+\) onto the irrep labelled by \(+\).

Let $k\in\Ring$. 
Let $\NullVector_k$ denotes the $k\times k$ matrix with every entry equal to 0.
Similarly, $\mathbb{I}_k$ denotes the $k\times k$ identity matrix.
\begin{lemma}\label{lemma:projector-symmetric}
We show that the sum of PL representations of the $\thegroup$ 
is diagonal with~$d$ ones first and then $d^2-d$ zeroes last;
i.e.,
 \begin{equation}
 \sum_{\CyclicElement \in \Ring_{\Order(\QuditDim)}^{ \QuditDim}}
 \RepPlRep\left(\RepCyclic(\CyclicElement)\right)
 = 
 \begin{bmatrix}
 \mathbb{I}_\QuditDim & \\
 & \NullVector_{\QuditDim^2-\QuditDim}
\end{bmatrix}.
\end{equation}
\end{lemma}
\begin{proof}
By definition,
the entries of the PL representation are of the form
\begin{equation}
 \RepPlRep\left(\RepCyclic(\CyclicElement)\right)
 =
\trace(
W_i^{\dagger} \RepCyclic(\CyclicElement) W_j \RepCyclic(\CyclicElement)^{\dagger}
).
\end{equation}
Thus, if either \(W_i\) or \(W_j\) are diagonal, then 
the entry
\(\RepPlRep\left(\RepCyclic(\CyclicElement)\right)_{i,j}=
\delta_{i,j}\)---diagonal matrices commute.
For the case of both \(W_i\) and \(W_j\) non-diagonal, 
we just need to notice that \(X\) is associated with the permutation \((1 \cdots d)\).
Thus, \(X \RepCyclic(\CyclicElement) X^{\dagger}\) permutes every diagonal entry of
\(\RepCyclic(\CyclicElement)\).
Therefore, by summing over \(\CyclicElement\) (see proof of
Lemma~\ref{lemma:sum-over-alpha}),
we obtain
 \begin{equation}
\sum_{\CyclicElement}
\trace(
W_i^{\dagger} \RepCyclic(\CyclicElement) W_j \RepCyclic(\CyclicElement)^{\dagger}
) 
=
\sum_{\CyclicElement}\RepPlRep(\RepCyclic(\CyclicElement))_{i,j}
= 0.
\end{equation}
\end{proof}

\begin{lemma}\label{claim:Projector}
  We show that the weighted sum, by the character of the standard irrep of
  \(S_\qudim\),
  is a projector onto the subspace \(\Irrep_0\); that is,
\begin{equation}\label{eq:Projector}
 \sum_{\Permutation,\CyclicElement\in\thegroup}
 \Character_\labelirrepstd(\Permutation) 
 \RepPlRep(\RepSymm(\Permutation)\RepCyclic(\CyclicElement))
 =
 \begin{bmatrix}
 0 & & \\
 &* & \\
 & & \NullVector_{\QuditDim^2-\QuditDim}
 \end{bmatrix},
\end{equation}
where $*$ denotes a non-zero square $\QuditDim-1$-dimensional matrix.
\end{lemma}
\begin{proof}
 
We evaluate the sum in Eq.~\eqref{eq:Projector} in parts: first we sum over~\(\CyclicElement\) and then over~\(\Permutation\).
The first sum is evaluated 
using Lemma~\ref{lemma:projector-symmetric} in Eq.~\eqref{eq:Projector}.
The result of the sum is
\begin{equation}
 \sum_{\Permutation,\CyclicElement}
 \Character_\labelirrepstd(\Permutation) \RepPlRep(\RepSymm(\Permutation)\RepCyclic(\CyclicElement))
  =
 \sum_{\Permutation} \Character_\labelirrepstd(\Permutation) 
 \RepPlRep\left(\RepSymm(\Permutation)\right)
 (\mathbb{I}_\QuditDim\oplus \NullVector_{\QuditDim^2-\QuditDim}).
\end{equation}
Because 
the standard irrep of \(S_\QuditDim\)
is not the trivial irrep,
the orthogonality of the characters of inequivalent irreps
leads to
\begin{equation}
 \sum_{\Permutation,\CyclicElement}
 \Character_\labelirrepstd(\Permutation) 
 \RepPlRep(\RepSymm(\Permutation)\RepCyclic(\CyclicElement))
 =
 \begin{bmatrix}
 0 & & \\
 &* & \\
 & & \NullVector_{\QuditDim^2-\QuditDim}
 \end{bmatrix},
\end{equation}
where $*$ denotes a non-zero square $\QuditDim-1$-dimensional matrix.
\end{proof}

From Lemma~\ref{claim:Projector} we obtain the following important result
for the estimation of the parameters of the AGF.
Let~$k,l\in\integers$.
Let $v=(v_1,\ldots, v_k)$ be a vector with $k$ entries. 
Then $v^{\oplus 2}$ denotes the vector $v^{\oplus 2} = (v_1,\ldots, v_k,
v_1,\ldots,v_k)$.
Similarly, 
we denote by $v^{\oplus l}$ the concatenation of~$l$ copies of $v$.
\grotten*
\begin{proof}
Notice that the PL vectorisation of $\dyad{0}$ and
$H\dyad{0}H^{\dagger}$ are
\begin{align}
 \kket{+}
 &= (1/\sqrt{\QuditDim},\NullVector_{\QuditDim-1})^{\oplus \QuditDim},\\
\kket{0}
 &= 
 (1/\sqrt{\QuditDim})\mathbb{I}_\QuditDim \oplus \NullVector_{\QuditDim^2-\QuditDim}.
\end{align}
Then multiplying by the projector \eqref{eq:Projector} we obtain $\NullVector$
for $\kket{+}$
and non-zero for~\(\kket{0}\).
\end{proof}

\end{document}